\documentclass[aps,prd,reprint,twocolumn,preprintnumbers,floatfix,nofootinbib]{revtex4-1}
\usepackage[utf8]{inputenc}
\pdfoutput=1
\usepackage{graphicx}
\usepackage{amsmath}
\usepackage{amsfonts}
\usepackage{amssymb}
\usepackage{xcolor,soul}
\usepackage{epstopdf}
\usepackage{xcolor}
\usepackage{float}
\usepackage{subfigure}

\usepackage{relsize}
\usepackage{graphics}
\usepackage{hyperref}
\usepackage{mathrsfs}
\usepackage{amssymb}
\usepackage{booktabs}
\usepackage[normalem]{ulem}
\usepackage{amsthm}
\usepackage{cancel}
\usepackage{fontawesome}   
\usepackage{tabularx,ragged2e} 

\hypersetup{
     colorlinks   = true,
     citecolor    = red,
     linkcolor    = blue,
     urlcolor     = blue,
}
\usepackage{physics}  
\newcommand{\bea}{\begin{aligned}}
\newcommand{\eea}{\end{aligned}}
\def\beq{\begin{equation}}
\def\eeq{\end{equation}}
\def\beqa{\begin{eqnarray}}
\def\eeqa{\end{eqnarray}}
\def\be{\begin{equation}}
\def\ee{\end{equation}}

\def\bse{\begin{subequations}}
\def\ese{\end{subequations}}

\def\trh{$T_{\mathrm{RH}}~$}
\def\mrm{\mathrm}
\def\tin{t_{\rm in}}
\def\ae{a_{\mathrm{end}}}
\def\arh{a_{\mathrm{RH}}}
\def\abh{a_{\mathrm{BH}}}
\def\arbh{a_R^{\mathrm{BH}}}
\def\rhorh{\rho_{\mathrm{RH}}}
\def\trh{T_{\mathrm{RH}}}
\def\bea{\begin{eqnarray}}
\def\eea{\end{eqnarray}}
\def\rhoe{\rho_{\mathrm{end}}}
\def\ai{a_{\mathrm{in}}}
\def\aev{a_{\mathrm{ev}}}
\def\Min{M_{\mathrm{in}}}
\def\BH{\mathrm{BH}}

\newcommand{\red}{\color{red}}
\newcommand{\blue}{\color{blue}}

\newcommand{\magenta}{\color{magenta}}

\graphicspath{{./figs/}}
\keywords{Primordial black holes , Reheating, Superradiance, Dark matter, Hawking radiation.}

\usepackage{pgfplots}
\pgfplotsset{compat=1.17}

\begin{document}




\preprint{}
\preprint{}

\vspace*{1mm}

\title{Primordial Black Hole Reheating}

\author{Md Riajul Haque$^{a}$}
\email{riaj.0009@gmail.com}

\author{Essodjolo Kpatcha$^{b,c}$}
\email{kpatcha@ijclab.in2p3.fr}

\author{Debaprasad Maity$^{d}$}
\email{debu@iitg.ac.in}

\author{Yann Mambrini$^{b}$}
\email{yann.mambrini@ijclab.in2p3.fr}

\vspace{0.1cm}

 \affiliation{
${}^a$
Centre for Strings, Gravitation and Cosmology,
Department of Physics, Indian Institute of Technology Madras, 
Chennai~600036, India
}

\affiliation{
${}^b$ Universit\'e Paris-Saclay, CNRS/IN2P3, IJCLab, 91405 Orsay, France
 }

\affiliation{${}^c$ Departamento de F\'{\i}sica Te\'{o}rica, Universidad Aut\'{o}noma de Madrid (UAM),
Campus de Cantoblanco, 28049 Madrid, Spain}

\affiliation{
${}^d$
Department of Physics, Indian Institute of Technology Guwahati, Guwahati, Assam, India
}

\begin{abstract} 
Post-inflationary reheating phase is usually said to be solely governed by the decay of coherently oscillating inflaton into radiation. In this submission, we explore a new avenue toward reheating through the evaporation of primordial black holes (PBHs). After the inflation, if PBHs form, depending on its initial mass, abundance, and inflaton coupling with the radiation, we found two physically distinct possibilities of reheating the universe. In one possibility, the thermal bath is solely obtained from the decay of PBHs while inflaton plays the role of dominant energy component in the entire process. In the other possibility, we found that PBHs itself dominate the total energy budget of the Universe during the course of evolution, and then its subsequent evaporation leads to radiation dominated universe. Furthermore, we analyze the impact of both monochromatic and extended PBH mass functions and estimate the detailed parameter ranges for which those distinct reheating histories are realized.


\end{abstract}

\maketitle

\tableofcontents

\section{Introduction}

Reheating is believed to be the most important phase of the early Universe, which successfully connects the super-cooled end of inflation phase with the standard hot Universe \cite{Nanopoulos:1983up}. Any observable imprints of this phase in the present Universe would be exciting due to its direct connection with beyond-standard model physics of cosmology and particle physics. With the advent of increasingly sophisticated experiments, reheating phase could be assumed as a perfect cosmological laboratory operating within a wide range of energy scales from the MeV to $10^{16}$ GeV. Over the years, attempts have been made both from particle phenomenology and cosmology to look for observables that can carry the interesting imprints of this phase. However, understanding the reheating mechanism is believed to be incomplete. The most common scenario advocates a homogeneous field, the inflaton, 
transferring its energy in the form of relativistic particles. 
This process can be non-perturbative \cite{Dolgov:1982th,Kofman:1997yn,Garcia:2021iag,Lebedev:2023zgw} 
or perturbative \cite{    
   Haque:2020zco,Garcia:2020wiy}
depending upon its coupling to the Standard Model (SM). 

Therefore, the reheating process is usually considered model-dependent, making it difficult to identify any observable that can encode the reheating histories. However, it has been shown recently that the gravitational interaction between the inflaton and the SM can be sufficient to reheat the Universe \cite{Mambrini:2021zpp,Haque:2022kez,Clery:2021bwz,Clery:2022wib,Haque:2023yra} without invoking additional couplings. Such gravitational reheating scenario usually predicts low reheating temperature with steep inflaton potential and are tightly constrained by the excessive production of high-frequency gravitational waves during BBN \cite{Haque:2021dha,Barman:2022qgt,Chakraborty:2023ocr,Barman:2023ktz}.

In this paper, we investigate another universal reheating mechanism where the radiation bath can be produced through the evaporation of  primordial black holes (PBHs). The formation of PBHs in the early Universe has been the subject of intensive investigation in recent times. 
If the amplitude of the local density fluctuation is strong enough, above
a critical value $\delta_c$
($\frac{\delta \rho}{\rho} \gtrsim \delta_c \sim 1$), PBHs can be shown to be produced by gravitational collapse. Several mechanisms for generating such a high local density fluctuation have been investigated, considering different physically motivated scenarios in the literature: quantum fluctuation generated during inflation through single field \cite{PhysRevD.48.543,PhysRevD.50.7173,Yokoyama:1998pt,Garcia-Bellido:2017mdw}, multi-fields \cite{Randall:1995dj,Garcia-Bellido:1996mdl}, collapse of cosmic sting loops during the radiation dominated Universe \cite{MacGibbon:1990zk,Jenkins:2020ctp,Helfer:2018qgv,Matsuda:2005ez,Lake:2009nq}, collapse of domain walls \cite{Rubin:2000dq,Rubin:2001yw}, bubble collision during phase transition \cite{KodamaPTP1979}. Instead of discussing the mechanism, we investigate the physical effects of PBHs once they are formed during reheating. In this paper, we particularly study the effect of evaporating PBHs on the reheating dynamics and investigate the possibility of getting a radiation-dominated Universe purely from PBH's evaporation. 


Indeed, if PBHs form during reheating, they can store a reasonable fraction of the total energy under the form of matter. 
Their density $\rho_{\rm BH}$ being less
affected by the dilution factor $a$ ($\rho_{\rm BH}\propto a^{-3}$), 
the PBH population can even dominate the energy budget of the Universe over the inflaton field. This happens, for instance, in the case of quartic potential $V(\phi) \sim \phi^4$
where the inflaton $\phi$, behaves like a radiation field
($\rho_\phi \propto a^{-4}$). Such phenomena of PBHs domination would be even easier to achieve for potentials $V(\phi)\sim \phi^n$ with $n>4$. 
Once the PBHs decay, they would release the amount of energy stored under the form of radiation, completing the reheating process.

As we will show, it is important to note that the PBHs {\it does not have} to dominate
over the inflaton density to affect the reheating. 
Even if they remain subdominant, the continuous entropy injection
through their decay can notably change the reheating process,
especially for low inflaton couplings to the particles in the plasma. 
Indeed,
the temperature of the thermal bath could sensibly increase due to
the fact that PBHs decay can easily generate thermal particles 
in much greater amounts than the inflaton decay itself.

The paper is organized as follows. After reminding the standard lore of reheating through the inflaton, we study in section III the evolution of PBHs, from their formation to their evaporation, in an expanding Universe. In section IV, we propose the possibility of completing the reheating through the decay of monochromatic primordial black holes formed during reheating. We generalize our analysis to extended mass distribution in section IV before concluding.


\section{Standard Reheating } 
\label{standard-reheating}

An important feature of inflationary models is the possibility of reheating the Universe after the inflation, leading to 
a radiation dominated epoch.
Inflaton reheating refers to the process by which the energy of the inflaton field, which powered the inflationary expansion of the Universe, is transferred to other particles in the Universe. This transfer of energy occurs at the end of the inflationary period and is considered to have created the conditions necessary for the formation of 
primordial nuclei and structures in the Universe. 
The transfer of energy from the inflaton to other particles is thought to have been accomplished through a variety of mechanisms, such as the decay of the inflaton into other particles or the production of particles through the interaction of the inflaton with other fields~\cite{Abbott:1982hn,Dolgov:1982th,Nanopoulos:1983up,Garcia:2020wiy}.

In our study, we assume that the reheating is not instantaneous, that is, a scenario in which the transfer of energy from the inflaton field to other particles at the end of inflation occurs over a longer period of time, rather than instantaneously. Note that there has been many works which have taken into account non-instantaneous reheating scenario, see for example Refs.~ \cite{Giudice:2000ex,Garcia:2017tuj,Dudas:2017rpa,Chen:2017kvz,Garcia:2020eof,Bernal:2020gzm,Co:2020xaf,Garcia:2020wiy}. 

 In the standard scenario, the evolution of the inflaton ($\rho_\phi$) and radiation ($\rho_R$) energy densities simply follow the set of Boltzmann equations
\beqa
 \dot{\rho_\phi}+3H(1+w_\phi)\rho_\phi &=& -\Gamma_\phi\rho_\phi\,(1+w_\phi) ,  \label{Eq:rhophi}\,\\
\dot{\rho}_R+4H\rho_R &=& \Gamma_{\phi}\rho_{\phi}(1+w_\phi)\, \label{rad}\label{Eq:rhor}\,,
\label{Eq:eqrhorwobh}
\eeqa
where $w_{\phi}$ is the equation of state for $\phi$, and 
is given by~\cite{Garcia:2020wiy}:
\beqa \label{eq:of:state}
w_{\phi} &=& \dfrac{n-2}{n+2}\, , 
\eeqa

\noindent
for a potential of the form $V(\phi)=\lambda M_P^4  ({\phi}/{M_P})^n$.
$\Gamma_\phi$ represents the decay or annihilation rate of the inflaton, which depends on the reheating process considered. $H$ is the Hubble parameter, and $M_P = 1/\sqrt{8\pi G} \simeq 2.435 \times 10^{18}$ GeV is the reduced Planck mass. The equations (\ref{Eq:rhophi}) and (\ref{Eq:rhor}), together with the Friedmann equation
\beq
\rho_{\phi}+\rho_{R} \;=\; 3H^2 M_P^2\,,
\label{hub}
\eeq 
allow to simultaneously solve for $\rho_\phi$ and $\rho_R$~\cite{Garcia:2020wiy}. It follows that the energy density of the inflaton and the radiation can be expressed in terms of the normalized scale factor $a/\ae$, $\ae$ being the scale factor at the end of inflation. We obtain
\beq
\rho_\phi(a) = \rhoe\left(\frac{\ae}{a}\right)^{3(1+w_\phi)} = \rho_{\rm end}\, \left(\frac{a}{\ae}\right)^{-\frac{6n}{n+2}} \, ,
\label{Eq:rhophiwobh}
\eeq

\noindent
and for $\rho_R$, supposing a coupling of the type $y_{\phi} \phi \bar f f$
between the inflaton and fermions (see appendix \ref{appendixA} for details),
\bea
\rho_R(a) &=& \frac{y_\phi^2}{8 \pi} \lambda^{\frac{1}{n}}  \alpha_n
\left(\frac{\rhoe}{M_P^4}\right)^{1-\frac{1}{n}} \left(\frac{a}{\ae}\right)^{-4}
\label{Eq:rhorwobh}
\\
&&
\times
\left[\left(\frac{a}{\ae}\right)^{2\frac{7-n}{n+2}}-1\right]
= \alpha_T T^4 \,,
\nonumber
\eea
where $\alpha_T = {g_T \pi^2}/{30}$ with $g_T$ is the number of relativistic degrees of freedom at temperature $T$ (106.75 for the Standard Model) and
\beq
\alpha_n=\frac{\sqrt{3n^3(n-1)}}{7-n}M_P^4\,.
\eeq
As we pointed out earlier, there are multiple possibilities for the reheating processes, from decay to bosonic states, scattering, or gravitational production.
Deferring the detailed study for our future work, in this paper we consider inflaton decaying into Fermions through $y_{\phi} \phi \bar f f$ interaction. 
The reheating is assumed to be completed at a scale $\arh$,
when $\rho_\phi(\arh)=\rho_R(\arh)=\rhorh$. 
Indeed, comparing Eqs.(\ref{Eq:rhophiwobh}) and (\ref{Eq:rhorwobh}), we see
that for $a \gg \ae$, ${\rho_\phi}/{\rho_R}\propto \left({a}/{\ae}\right)^{-\frac{6}{n+2}}$ for $n< 7$, and ${\rho_\phi}/{\rho_R}\propto \left({a}/{\ae}\right)^{-\frac{2n-8}{n+2}}$ for $n>7$, both of them decreasing with the scale factor $a$. This condition can also be seen to be true for the special case $n=7$. The condition mentioned above immediately suggests, therefore, that there exists a value $a=\arh$ for which $\rho_\phi=\rho_R$. Considering $n<7$, this happens for 
\bea
&&
\left(\frac{\arh}{\ae}\right)^{-\frac{6}{n+2}}
=\frac{y_\phi^2}{8 \pi}
\left(\frac{\alpha_n}{M_P^4} \right) \left(\frac{{\lambda} M_P^4}{\rhoe}\right)^\frac{1}{n}
\label{Eq:arhnobh}
\\
&&
\Rightarrow \rhorh=\rhoe\left(\frac{\arh}{\ae}\right)^{-\frac{6n}{n+2}}
=\left(\frac{y_\phi^2}{8 \pi}\right)^n\left(\frac{\alpha_n}{M_P^4}\right)^n \lambda M_P^4
\nonumber
\eea
\noindent
or
\beq
\trh^{n<7}\simeq \frac{4.3\times 10^{15}}{2.3(2.5 \times 10^9)^\frac{n}{4}}
\left[\frac{\alpha_n}{M_P^4}\right]^{\frac{n}{4}}
\left[\frac{y_{\phi}}{10^{-4}}\right]^{\frac{n}{2}}\left[\frac{\lambda}{10^{-11}}\right]^{\frac{1}{4}} .
\label{Eq:trhwobh}
\eeq
Similar analysis for $n>7$ gives the following expression for the reheating temperature and scale factor at the end of the reheating 
\bea
\trh^{n>7} &\simeq& \frac{(4.3\times 10^{15})^{\frac{3}{n-4}}}{2.3(2.5 \times 10^9)^\frac{3n}{4n-16}}
\left[\frac{-\alpha_n}{M_P^4}\right]^{\frac{3n}{4n-16}}
\left[\frac{y_{\phi}}{10^{-4}}\right]^{\frac{3n}{2n-8}} \nonumber
\\
&&\times \left(\frac{\lambda}{10^{-11}}\right)^{\frac{3}{4n-16}} \left(\rho_{\rm end}\right)^{\frac{n-7}{4n-16}} 
\label{Eq:trhwobhn>7}
\eea
with
\bea\label{Eq:an>7}
\left(\frac{\arh}{\ae}\right)^{\frac{2(4-n)}{n+2}}=\frac{y_\phi^2}{8 \pi}
\left(\frac{-\alpha_n}{M_P^4} \right) \left(\frac{{\lambda} M_P^4}{\rhoe}\right)^\frac{1}{n}
\eea

\begin{figure}[!ht]
\centering
\includegraphics[width=\columnwidth]{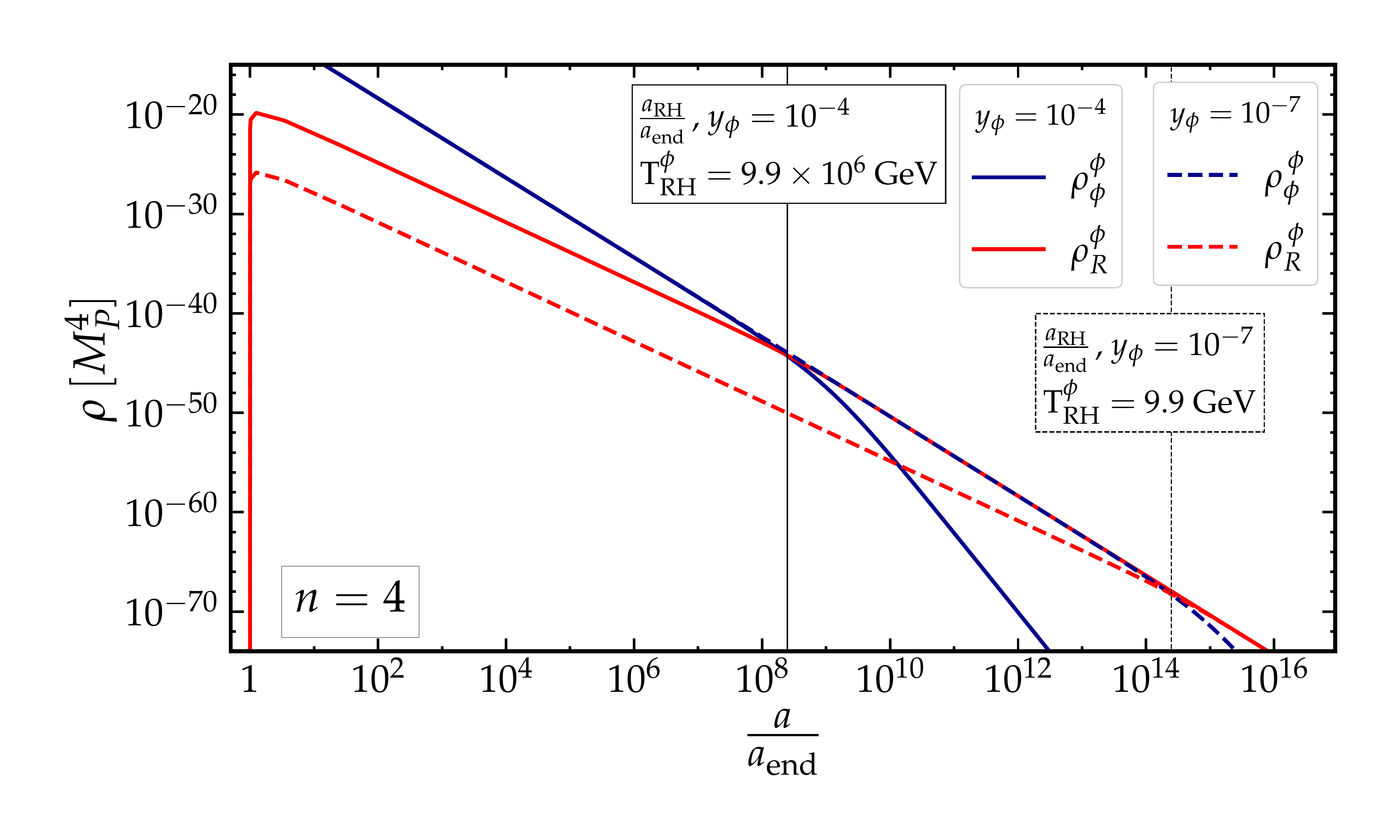}
\caption{\em \small Evolution of the inflaton density, $\rho_{\phi}$,
and the radiation density $\rho_R$ as function of $a/\ae$ for different values of the Yukawa coupling, $y_\phi=10^{-7}$ and $10^{-4}$.}
\label{Fig:reheatingwobh}
\end{figure}

In the above expression, $\trh$ is expressed in GeV, and we took $g_T = 106.75$. We show in Fig.(\ref{Fig:reheatingwobh}) the corresponding evaluations
for $\rho_\phi$ and $\rho_R$ as a function of $a/\ae$ for $n=4$. We considered two different values of the Yukawa couplings, $y_{\phi}=10^{-7}$ and $y_\phi=10^{-4}$, giving rise to reheating temperature $\trh\simeq 10$ GeV and $10^7$ GeV, respectively, for\footnote{See details in the appendix for the values of $\lambda$ and $\rhoe$ we used.} $\lambda = 5\times 10^{-11}$, and $\rhoe=1.45\times 10^{63}~\mrm{GeV^4}$. 

The presence of PBHs is expected to affect significantly the vanilla scenario discussed above, adding a new matter component in the primordial plasma.  
As we will see, there even exists the possibility that the PBHs and its decay products may either dominate the inflaton energy density or the primordial plasma. The reheating is then completed through their decay. As a consequence, the reheating temperature can be drastically different from the one obtained  in Eqs.(\ref{Eq:trhwobh}) and (\ref{Eq:trhwobhn>7}).

\section{PBH evolution}
\label{section-monochromatic}

\subsection{Mass function}

PBHs could have been produced during the early Universe due to various mechanisms. The common point for all of them is the collapse of a spatial region with large primordial energy density fluctuations due to gravitational pull. One possibility is that PBHs formed in a relatively short period of time,
in regions where $\frac{\delta \rho}{\rho} \gtrsim \delta_c \sim 1$. This could have happened, for example, during a phase transition, where the properties of matter changed rapidly, leading to the collapse of large regions into black holes. In this scenario, the mass distribution of primordial black holes would be concentrated, sharply peaked, or monochromatic around a specific mass. The mass distribution is then just a delta function:
\beqa
\label{monochromatic_mass_function}
f_{\rm PBH}(M) = \delta(M - M_{\rm in})\,.
\eeqa
The initial mass of such PBHs $M_{\rm in}$ is assumed to be a fraction of the Hubble mass $M_H$ at the time of formation, $\tin$~\cite{Harada:2013epa,Sasaki:2018dmp,Villanueva-Domingo:2021spv}.
In fact, the PBHs are supposed to be formed almost instantaneously. The initial mass 
is then naturally of the order of the energy embedded in the horizon at the formation time, and can be written
\beqa
M_{\rm in} = \gamma\,M_H=\gamma \frac{4\pi}{3} \dfrac{\rho(t_{\rm in})}{H^3(\tin)} =4\pi \gamma\frac{M_P^2}{H(\tin)}\,,
\label{Eq:pbh-m-in}
\eeqa

\noindent
where, $\gamma=w^{3/2}$ is the efficiency of the collapse \cite{Carr:1974nx}. 
Note that for radiation dominated background, $w={1}/{3}$, the value of $\gamma$ assumes $\sim 0.2$, whereas
for $w=0$ the formation mechanism is more involved. Nonetheless, for all the cases we considered $M_{\rm in}$ as a free parameter.
Moreover, certain mechanisms suggest that PBHs may exhibit an extended mass function, i.e., a distribution of masses~(see Ref.~\cite{Carr:2020xqk,Carr:2021bzv} and references therein). The existence of an extended mass distribution is attributed to the formation of PBHs from density perturbations of varying scales. Specifically, smaller perturbations would have given rise to smaller black holes, while larger perturbations would have led to the formation of larger black holes.
There exists also the possibility for a more complex spectrum which we will analyze in the following section.

Note also that the initial PBH mass is bounded by the size of the horizon at the end of inflation,
\beq
\Min \gtrsim H_{\rm end}^{-3} \rho_{\rm end} \sim \frac{M_P^3}{\sqrt{\rho_{\rm end}}}\simeq 1\mrm{g} =M_{ \rm min}
\eeq

\noindent where we took $\rho_{\rm end}^{1/4} \sim 10^{15}$ GeV. We will also consider black holes decaying before BBN to avoid perturbations due to entropy injection from evaporating PBHs. Indeed, in a seminal 
paper\cite{Hawking:1974rv}, Stephen Hawking opened the possibility for black holes to evaporate, into a radiation corresponding to temperature $T_{\rm BH} \sim {M_P^2}/{M_{\rm BH}}$. As a consequence, the black hole decays, and its mass varies with time as \cite{Hawking:1975vcx}
\beqa \label{massev}
\frac{dM_{\rm BH}}{dt} = - \epsilon \frac{M^4_P}{M_{\rm BH}^2} \, ,
\label{Eq:evaporation}
\eeqa

\noindent
where $\epsilon = \pi g_{\rm BH}/{480}$, with $ g_{\rm BH}$  the number of degrees of freedom below $T_{\rm BH}$\footnote{Notice that in the expression of $\epsilon$ we assume the gray-body factor to be $\mathcal{G} = 1$. Nevertheless, a proper treatment would consider $\mathcal{G} \approx 3.8$ (see Ref. \cite{Masina:2020xhk}).}.
Solving Eq.(\ref{Eq:evaporation}) one obtains the evaporation time of PBH, $t_{\rm ev}$ :
\beq
t_{\rm ev} \simeq 1~ \mrm{s} ~\left(\frac{\Min}{10^8~\mrm{g}}\right)^3.
\eeq

\noindent
The typical time of BBN being of order one second, we will restrict our analysis within the following mass range of PBHs,   
\beq
1 \mrm{g} \lesssim \Min \lesssim 10^8 \mrm{g}.
\eeq

\subsection{PBH energy density}
We consider that a fraction $\beta$ of the total energy falls into black holes,
\be
\beta=\frac{\rho_{\rm BH}(t_{\rm in})}{\rho_{\rm tot} (t_{\rm in})}\,,
\ee

\noindent
where $\rho_{\rm tot} = \rho_\phi + \rho_{R}$ is the total energy density.
$\beta$ can be restricted by imposing constraints from the induced gravitational waves (GWs), generated at second order in perturbation theory, sourced by the density fluctuation due to the inhomogeneities of the PBH distribution before it evaporates. The produced GWs energy density either can overtake the background energy density or severely impact the Big-Bang Nucleosynthesis (BBN) processes \cite{Inomata:2020lmk,Hooper:2020evu,Papanikolaou:2020qtd,Domenech:2020ssp}. For instance, in Ref.~\cite{Papanikolaou:2020qtd}, an upper limit on the value of 
\beq\beta < 10^{-4} \left(M_{\rm in}/10^{9} \, {\rm g} \right)^{-1/4}
\nonumber
\eeq
has been derived which avoid backreaction problem. However, a stronger upper limit derived in Ref.~\cite{Domenech:2020ssp} asserts that the dominant contribution to GWs arises from the sudden evaporation of PBHs in the PBH domination regime. Specifically, this upper bound on $\beta$ is obtained by demanding that the amount of generated GWs is not in conflict with the BBN constraints on the effective number of relativistic species and is expressed as:
\begin{multline} \label{upper-bound-beta}
\beta < 1.1 \times 10^{-6} \left(\frac{\gamma}{0.2} \right)^{-1/2} \left(\frac{g_{\rm BH}}{108} \right)^{17/48} \\
\left(\frac{g_{\rm ev}}{106.75} \right)^{1/16} \left(\frac{M_{\rm in}}{10^{4}\, {\rm g}} \right)^{-17/24} \, .
\end{multline}
where $g_{\rm ev}$ is the number of degrees of freedom at the evaporation time $t_{\rm ev}$. We have applied the above-mentioned upper limit for $\beta$ throughout our analysis.

The evolution of the energy density of the primordial black hole before its evaporation takes the following form,
\beqa
\dot{\rho}_{\rm BH}+3H\rho_{\rm BH}=\frac{\rho_{\rm BH}}{M_{\rm BH}}\frac{dM_{\rm BH}}{dt} \,\theta(t-t_{\rm in})\,\theta(t_{\rm ev}-t)\,,
 \label{Eq:pbhevolution}
\eeqa
where the $\theta$-function is the Heaviside function and $t_{\rm in}$ ($t_{\rm ev}$) is the time associated with the formation (evaporation) point. The PBH energy density is obtained
by solving Eq.(\ref{Eq:pbhevolution}) while respecting
(\ref{Eq:evaporation}). In a Universe
whose expansion is dominated by a fluid with an equation of state $P=w \rho$, solving Eq.(\ref{Eq:evaporation}) gives
\bea
&&
M_{\rm BH}^3(a)=\Min^3+\frac{2 \epsilon M_P^2\Min}{4 \pi \gamma(1+w)}
\left[1-\left(\frac{a}{\ai}\right)^{\frac{3}{2}(1+w)}\right]
\nonumber
\\
&&\simeq \Min^3\left[1-\frac{2 \sqrt{3} \epsilon}{1+w}\frac{M_P^5}{\Min^3\sqrt{\rho_{\rm end}}}\left(\frac{a}{\ae}\right)^{\frac{3}{2}(1+w)}\right]\,,
\label{Eq:mbh}
\eea

\noindent
where we supposed for the last approximation $a\gg \ai$.
We also used Eq.(\ref{Eq:pbh-m-in}) to write
\bea
\frac{\ai}{\ae}&=&\left(\frac{\Min \sqrt{\rhoe}}{4 \pi \gamma \sqrt{3}M_P^3}\right)^\frac{2}{3(1+w)}
\\
&\simeq&
\left[1.7\times 10^{-2}\left(\frac{\Min}{1~\mrm{g}}\right)\sqrt{\frac{\rhoe}{10^{60}}}\right]^{\frac{2}{3(1+w)}}
\,,
\nonumber
\eea

\noindent 
the unit being in GeV when not specified.
Notice that the evaporation time, or scale factor $\aev$
can also be deduced from Eq.(\ref{Eq:mbh}). Asking
for $M(\aev)=0$, we have
\bea
\frac{\aev}{\ae}&=&
\left[\frac{(1+w)}{2 \sqrt{3}\epsilon}
\frac{\Min^3\sqrt{\rhoe}}{M_P^5}\right]^{\frac{2}{3(1+w)}}
\label{Eq:aev}
\\
&\simeq&
\left[
4.5\times 10^8\left(\frac{1+w}{\epsilon}\right)
\left(\frac{\Min}{1~\mrm{g}}\right)^3\sqrt{\frac{\rhoe}{10^{60}}}
\right]^\frac{2}{3(1+w)}
\,.
\nonumber
\eea

\noindent
The effect of the black hole evaporation on the reheating process will then last from $\ai$ till $\aev$. It is also interesting to note that the dimensionless 
factor ${M_P^5}/{\Min^3\sqrt{\rhoe}}$ appearing
in Eqs.(\ref{Eq:mbh}) and (\ref{Eq:aev}) is very small, reaching a maximum
value of $\sim 10^{-12}$ for $\Min=1$ gram, the PBHs mass at the end of inflation, $\ae$. 
That justifies the approximation $\aev \gg \ae,  \ai$.
Depending upon the initial abundance of lighter PBHs such a long period of decay time with, ${\aev}/{\ae} > 10^{12}$ can give rise to different physically distinct reheating scenarios such as PBHs dominating the Universe before their complete evaporation, or PBHs evaporation unilaterally completing the reheating processes. In the subsequent sections, we will discuss various scenarios in detail.

Implementing Eq.(\ref{Eq:mbh}) in (\ref{Eq:pbhevolution}) and using Eq.(\ref{Eq:pbh-m-in}), we finally obtain
\bea
\rho_{\rm BH}(a)&=&\beta \rhoe 
\left(\frac{4 \pi \sqrt{3}\gamma M_P^3}{\Min\sqrt{\rhoe}}\right)^\frac{2w}{(1+w)}
\left(\frac{\ae}{a}\right)^3
\label{Eq:rhobh}
\\
&\times&
\left[1-\frac{2 \sqrt{3}\epsilon}{1+w}\frac{M_P^5}{\Min^3\sqrt{\rhoe}}\left(\frac{a}{\ae}\right)^{\frac{3}{2}(1+w)}\right]^\frac{1}{3}\,.
\nonumber
\eea

\begin{figure}[!ht]
\centering
\includegraphics[width=\columnwidth]{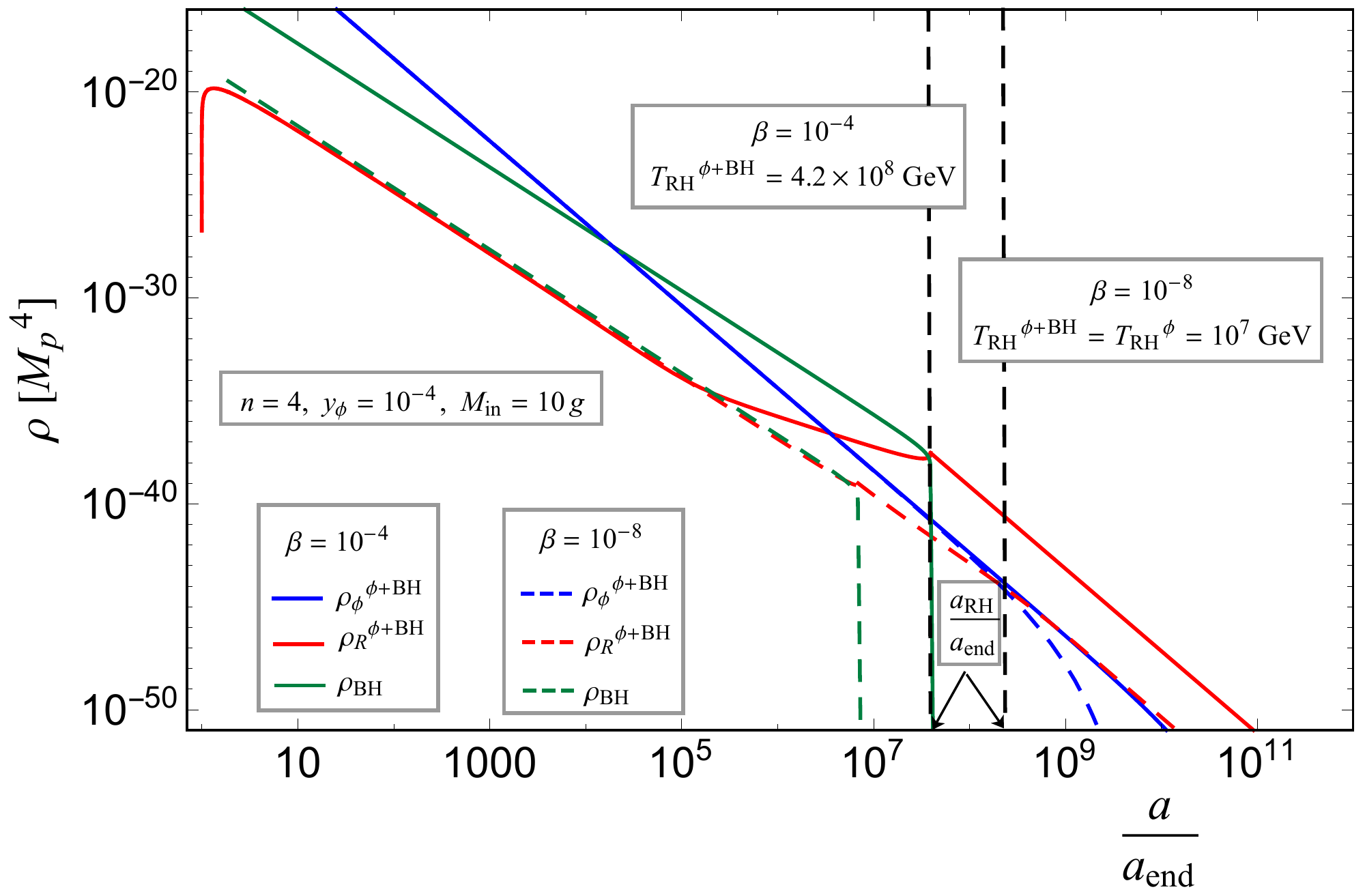}
\caption{\em \small Evolution of the energy densities $\rho_{\phi}$ (blue), $\rho_R$ (red) and $\rho_{BH}$ (green)
as function of $a/\ae$ for $n=4$, $y_\phi=10^{-4}$, $\beta=10^{-8}$ (dashed) and $10^{-4}$ (full). 
Note the shift in the PBHs lifetime if
they dominate the energy budget of the Universe before decaying.}
\label{Fig:reheatingwbh}
\end{figure}

\noindent
Note that, in the case of negligible evaporation
($a\ll a_{\rm ev}$) , the second term in the third bracket is always subdominant and we can recover the usual pressureless dust-like nature of PBHs energy density with $\rho_{\rm BH}\propto a^{-3}$, 
and proportional to the part of the energy density collapsing, $\beta$, and the efficiency of the collapse, $\gamma$. Furthermore, for $a=\aev$, 
we recover $\rho_{\rm BH}=0$, as expected. We illustrate in Fig.(\ref{Fig:reheatingwbh}) the evolution of $\rho_{\rm BH}$ as a function 
of ${a}/{\ae}$ for the same set of parameters as in Fig.(\ref{Fig:reheatingwobh}). To obtain the figure, we solved numerically the set of Eqs. (\ref{Eq:rhophi}) and (\ref{Eq:pbhevolution}), for two values of fraction $\beta=(10^{-8},10^{-4})$ and $\Min=10$ g. $V(\phi)$ being quartic, $n=4$ implies $w=w_\phi=\frac{1}{3}$. We clearly observe the $\rho_{\rm BH}\propto a^{-3}$ behavior as expected before the evaporation, which is almost instantaneous and happens for ${\aev}/{\ae}\simeq5\times 10^6$ in the case $\beta=10^{-8}$, in accordance with Eq.(\ref{Eq:aev}).
We also note that $\rho_{\rm BH}$ is proportional to $\beta$ as expected 
from Eq.(\ref{Eq:rhobh}).

\subsection{PBH domination?}

It can also be interesting to wonder if PBH can dominate the energy budget of the Universe {\it before} the end of the reheating process. One then needs to compute the time
$\abh$ when $\rho_\phi\sim\rho_{\rm BH}$. Indeed, for the PBHs behaving as dust before its
decay, $\rho_{\rm BH} \sim a^{-3}$, whereas the inflaton field follows $\rho_\phi\propto a^{-3(1+w_\phi)}$, there exist a point where $\rho_\phi=\rho_{\rm BH}$. Combining 
Eqs.(\ref{Eq:rhophiwobh}) and (\ref{Eq:rhobh}), one obtains
\bea \label{Eq:abhpbh}
\frac{\abh}{\ae} &=&\left(\frac{\Min \sqrt{\rhoe}}{4 \pi \gamma \sqrt{3}M_P^3}\right)^\frac{2}{3(1+w_\phi)}\beta^{-\frac{1}{3w_\phi}} \nonumber \\
&&\simeq \beta^{-\frac{1}{3w_\phi}}\left[\left(\frac{\Min}{112\mrm{g}}\right)\sqrt{\frac{\rho_{\rm end}}{10^{60}}}\left(\frac{0.2}{\gamma}\right)\right]^{\frac{2}{3(1+w_\phi)}}
\label{Eq:abh}
\eea
which gives, for $\beta=10^{-4}$,  $\Min=10$ g, $\rho_{\rm end}\sim 1.5\times 10^{63}$ GeV and $w_\phi=1/3$, 
${\abh}/{\ae}\sim 2\times10^4$,
values that we can recover in  Fig.(\ref{Fig:reheatingwbh}). However, domination of PBH does not occur for any value of $\beta$. There exists a critical value, denoted as $\beta_{\rm crit}^{\rm BH}$, above which PBHs dominate over the background energy density; in this scenario, the background is governed by inflaton. Indeed, this should happen before its total evaporation, in other words, $\abh < \aev$, or combining Eqs.(\ref{Eq:aev}) with (\ref{Eq:abh}), 
\begin{eqnarray}
\label{Eq:betamin}
\beta_{\rm crit}^{\rm BH} & = & \left(\frac{\epsilon}{(1+w_\phi)2 \pi \gamma}\right)^{\frac{2w_\phi}{1+w_\phi}} \left(\frac{M_P}{\Min}\right)^{\frac{4w_\phi}{1+w_\phi}} \,.
\end{eqnarray}
This corresponds to {$\beta \simeq 3\times 10^{-6}$ } for a quartic potential ($w_\phi=1/3$), {and $\Min = M_{\rm min} = 1$ g}.
On the other hand, for $\beta=10^{-8}$ Eq.(\ref{Eq:abh})
gives , $\abh/\ae\simeq 2\times10^{8}$, 
whereas $\aev\simeq 5 \times 10^{6}$, so there is no PBH domination, which is also what we observe in Fig.(\ref{Fig:reheatingwbh}).
In this case, the PBH population would {\it never} constitute the main component of
the Universe. Note that to get this particular scenario (PBH domination after inflation domination), $a_{\rm BH}<a_{\rm RH}$ and that will happen if the inflation-radiation coupling is less than some specific value $y_\phi^{\rm cst}$, which we defined later in Eq.(\ref{Eq:yukconstantTRH}) for $n<7$ and Eq.(\ref{Eq:yukconstantTRHng7}) for $n>7$. Otherwise, there will always be radiation domination after inflaton domination, and above some critical value of $\beta$, there is a possibility of PBH domination after radiation domination. One important point is to note that once we fixed $y_\phi<y_\phi^{\rm cst}$, $\beta_{\rm crit}^{\rm BH }$ is independent of the value of $y_\phi$ and determined from Eq.(\ref{Eq:betamin}) however for $y_\phi>y_\phi^{\rm cst}$, another critical value of $\beta$ for PBH domination is always a function of $y_\phi$ and the Eq.(\ref{Eq:betamin}) is not valid anymore. \\
The  domination of PBHs over the inflaton significantly affects the expansion rate $H=\sqrt{\rho_{\rm BH}/{3 M_P^2}}$,
and then the PBH lifetime itself. Indeed, the solution of (\ref{Eq:evaporation}) in a PBHs dominated Universe becomes
\beq
M_{\rm BH}^3(a) \simeq \Min^3-
\frac{2 \sqrt{3} \epsilon M_P^5}{\sqrt{\rho_\phi(\abh)}}\left(\frac{a}{\abh}\right)^\frac{3}{2}\,,
\eeq
where we supposed $M_{\rm BH}(\abh)\simeq \Min$ and $a \gg \abh$.
We then obtain the evaporation time
\beq \label{Eq:aevpbh}
M(\aev)=0~~\Rightarrow ~~\frac{\aev}{\abh}=\frac{\Min^2\rhoe^\frac{1}{3}}{(2 \sqrt{3}\epsilon M_P^5)^\frac{2}{3}}
\left(\frac{\ae}{\abh}\right)^{(1+w)}\,,
\eeq
where ${\ae}/{\abh}$ is given by (\ref{Eq:abh}).
If we take $\Min=10$ g and $\rhoe=1.46\times 10^{63}$, 
we find for $w=1/3$, 
${\aev}/{\ae}\sim 3\times 10^7$, corresponding to a little delay in  the PBH lifetime compared to the value 
$5\times 10^6$ that we obtained solving (\ref{Eq:rhobh})
where the inflaton was dominating the evolution of the Universe.
We also clearly see this shifting effect in the decay in Fig.(\ref{Fig:reheatingwbh}).
Note that it is not strictly speaking the {\it lifetime} which is changing,
but the corresponding scale factor due to a modification in the rate 
of expansion between an inflaton-dominated Universe and a PBH domination.\\
The PBHs evaporation produces SM particles which populate
the thermal bath. 
The evolution of the radiation energy density is then affected, $\rho_R$ receiving 
a new contribution from the decaying PBH. Eq.~(\ref{Eq:eqrhorwobh}) becomes:
\be
\dot{\rho}_R+4H\rho_R=\Gamma_{\phi}\rho_{\phi}(1+w_\phi)-\frac{\rho_{\rm BH}}{M_{\rm BH}}\frac{dM_{\rm BH}}{dt} \,\theta(t-t_{\rm in})\,\theta(t_{\rm ev}-t)\, .
\label{Eq:eqrhorwbh}
\ee

\noindent The dynamics of the system are determined by simultaneously solving Eq. (\ref{Eq:rhophi}), (\ref{Eq:pbhevolution}) and (\ref{Eq:eqrhorwbh}), together with the Friedmann equation
\beq
\rho_{\phi}+\rho_{R} + \rho_{\rm BH} \;=\; 3H^2 M_P^2\,.
\label{Eq:hub}
\eeq  

Different scenarios are expected depending on which component of the energy density dominates the Universe at subsequent epochs after the formation of PBHs. For instance, for small values of the Yukawa coupling $y_{\phi}$, the PBHs can regulate
the reheating process through entropy injection in such a way that there exists a lower bound on $y_{\phi}$ over which the inflaton decays {\it before} the PBH population.
Therefore, the Universe enters into PBH dominated phase which can drastically modify the reheating history. In the following section, we will study all the possible scenarios in detail step by step. 

\section{PBH reheating}

\subsection{generalities}

\begin{figure*}[t!]
	\includegraphics[height=12cm]{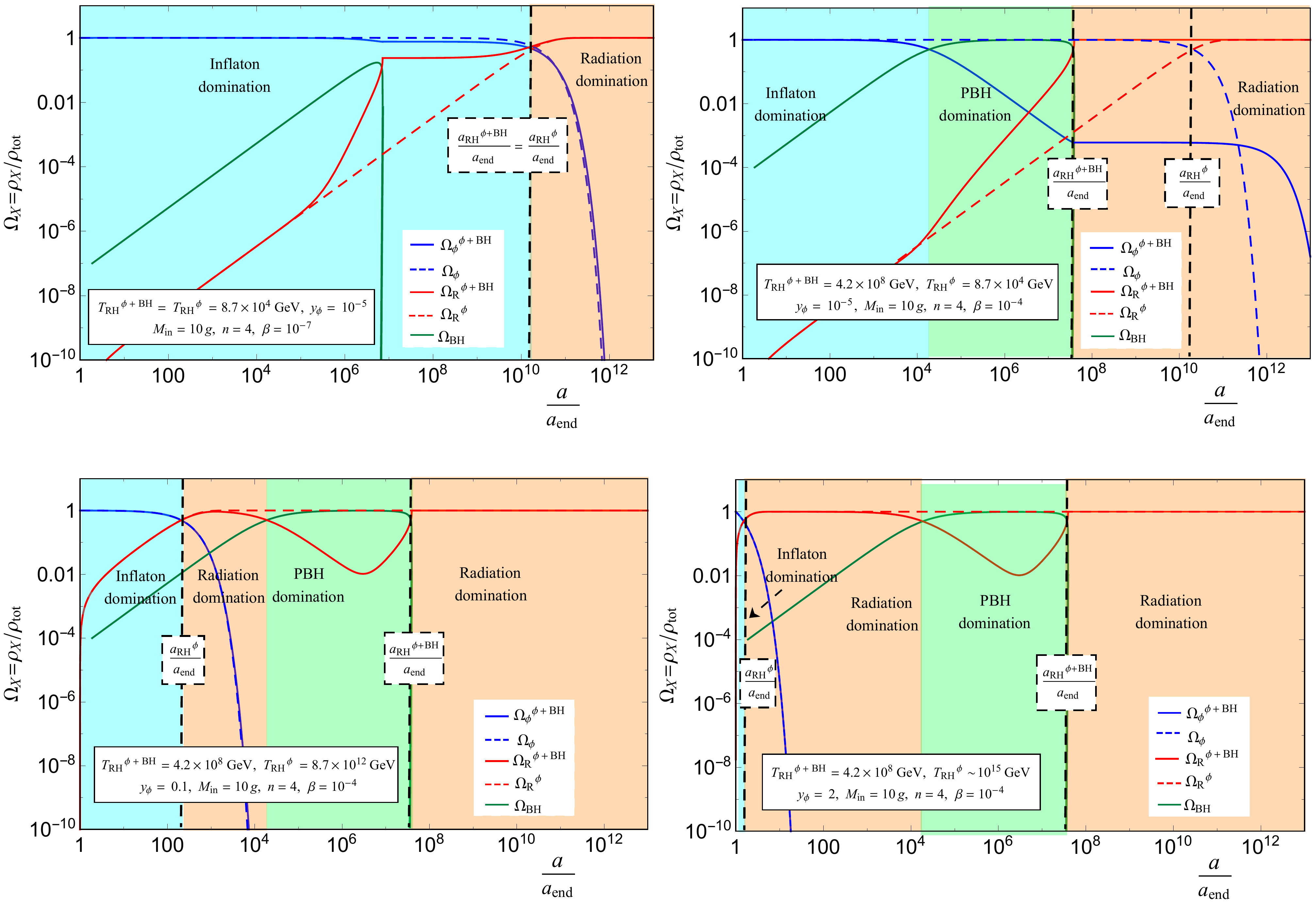}       
	\caption{\em Evolution of the normalized energy densities $\Omega_{X}=\frac{\rho_X}{3 M_{ p}^2\,H^2}$ as a function of scale factor for the different combination of $(y_\phi,\,M_{\rm in},\,\beta)$ with $n=4$. In the symbol of dimensionless energy densities, the $\phi+\rm BH$ and $\phi$ term indicates reheating dynamics with and without  black hole respectively. } 
	\label{Fig:reheatn4}
\end{figure*}
\begin{figure*}[t!]
	\includegraphics[height=6.05cm]{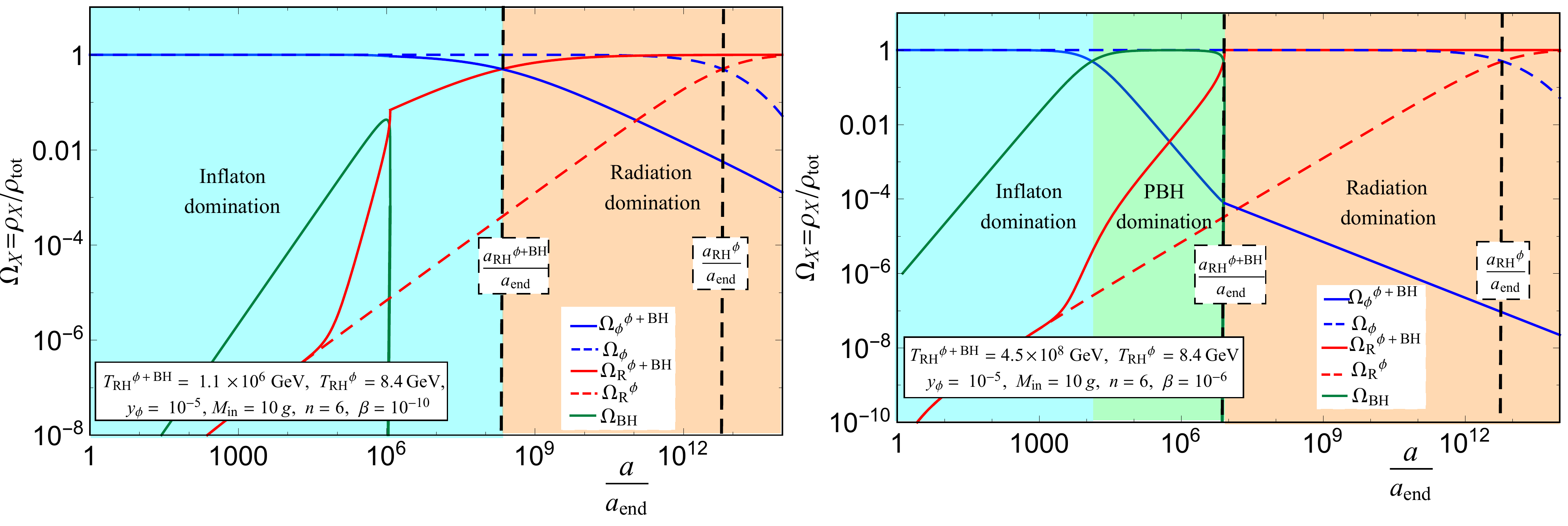}
 \includegraphics[height=6cm]{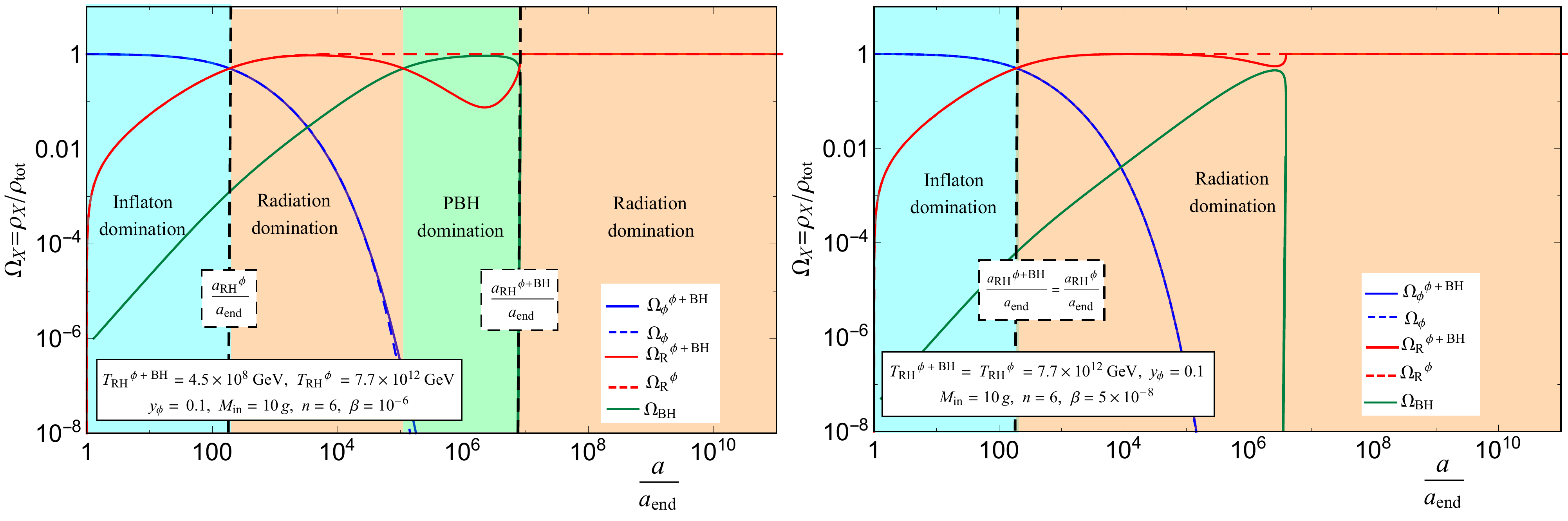}
	\caption{\em The description of this plot is same as Fig.\ref{Fig:reheatn4}. Here we have plotted for $n=6$.}
	\label{Fig:reheatn6}
\end{figure*}

Once the PBHs are produced, they can dominate the reheating process if
\beq
\Gamma_\phi \rho_\phi(1+w_\phi) < -\frac{\rho_{\rm BH}}{M_{\rm BH}} \frac{dM_{\rm BH}}{dt}\,,
\eeq

\noindent
corresponding to a scale factor $a=\arbh$
\bea
\left(\frac{\arbh}{\ae}\right)^{6w_\phi}
&\gtrsim& \frac{y_\phi^2}{\epsilon \beta}\frac{\lambda^{\frac{1-w_\phi}{2+2w_\phi}}\sqrt{n(n-1)}}{8 \pi (48 \pi^2 \gamma^2)^{\frac{1}{3+3w_\phi}}}
\label{Eq:arbh}
\\
&&
\times \left(\frac{\rhoe}{M_P^4}\right)^{\frac{2 w_\phi}{1+w_\phi}}\left(\frac{\Min}{M_P}\right)^{\frac{5 w_\phi+3}{1+w_\phi}}\,.
\nonumber
\eea

\noindent
where we combined Eqs.(\ref{Eq:rhophi}) and (\ref{Eq:rhobh}); considering $M_{\rm BH}\sim \Min$. 
Concerning the notation, $\arbh$, scale at which the PBHs dominates the {\it reheating process} 
should not be confused 
with $\abh$ from (\ref{Eq:abh}) which is the scale when the PBH population dominates the {\it energy density} of the Universe. The PBHs can indeed lead the reheating process even if they do not dominate over the inflaton density. Such PBHs-dominated reheating naturally predicts higher reheating temperature due to extra entropy injection, see Figs.(\ref{Fig:reheatn4})
and (\ref{Fig:reheatn6}), compared to the vanilla reheating scenario, as we will describe later.

From Eq.(\ref{Eq:arbh}) we can deduce that
the reheating process is driven by the PBH when
$\arbh/\ae=5\times 10^5$ in the case $n=4$ ($w=1/3$)
with $(y_\phi=10^{-5},\beta=10^{-7})$ and $\arbh/\ae=1.7\times 10^4$ for $(y_\phi=10^{-5},\beta=10^{-4})$. 
We summarize this behavior in  Figs.(\ref{Fig:reheatn4})
and (\ref{Fig:reheatn6}) in the case $n=4$ and $n=6$ respectively, for different values of $y_\phi$ and $\beta$
fixing $\Min$ to 10 grams. We recover the values of $\arbh/\ae$ for the two values of $\beta$ we just computed
analytically in Figs.(\ref{Fig:reheatn4}) top-left and top-right, where the change of slope in $\rho_R$ is obvious
at the corresponding values of $a=a_R^{\rm BH}$.

We can even recover the change in the slope of $\rho_R$
between the phase when the radiation is generated by the inflaton and the phase when it is driven by the decay of PBHs. Indeed, if PBHs become the source of radiation, Eq.(\ref{Eq:eqrhorwbh}) can be simplified by
\beq
\dot \rho_R + 4H\rho_R=\epsilon \rho_{\rm BH}\frac{M_P^4}{\Min^3}\,,
\label{Eq:rhorbh}
\eeq
where we supposed $M=\Min$ during the whole reheating process. The solution
of Eq.(\ref{Eq:rhorbh}) for $a \gg \arbh$ is then given by 
\beq
\rho_R^{\rm BH}(a)\simeq\rho_R(\arbh)\left(\frac{\arbh}{a}\right)^{\frac{3}{2}-\frac{3}{2}w}\,,
\label{Eq:rhorfrombh}
\eeq
where the upper index $\rm BH$ indicates the source of the radiation from the black hole and $w$
is the equation of state parameter of the field driving the expansion 
during the reheating ($w=w_\phi$ if the inflaton dominates, while $w=0$ if the PBHs dominate). We finally obtain when the 
inflaton dominates the energy budget
\beq
\rho_\phi\propto a^{-3(1+w_\phi)}\,,~~\rho_R^\phi\propto a^{-\frac{3}{2}(1+3w_\phi)}\,,~~
\rho_R^{\BH}\propto a^{-\frac{3}{2}(1-w_\phi)}\,.
\eeq
If the background is inflaton dominated, then the above equations give for $n=4$,
\beq
\frac{\rho^\phi_R}{\rho_\phi}\propto a \,, ~~~ \frac{\rho_R^{\rm BH}}{\rho_\phi}\propto a^3\,,
\eeq
and for $n=6$
\beq
\frac{\rho_R^\phi}{\rho_\phi}\propto a^{\frac{3}{4}}\,,
~~~\frac{\rho_R^{\rm BH}}{\rho_\phi}\propto a^{\frac{15}{4}}\,.
\eeq
This is exactly what we observe in Figs.(\ref{Fig:reheatn4})
and (\ref{Fig:reheatn6}) top-left. 
For larger values of $\beta$, the PBH population can dominate over the inflaton density, and we should consider $w=0$ in Eq.(\ref{Eq:rhorfrombh}),
which gives 
\beq
\frac{\rho_R^{\BH}}{\rho_{\BH}}\propto a^{\frac{3}{2}}\,,
\eeq
independently on $n$ of course, which is also what is observed in Figs. (\ref{Fig:reheatn4}) and (\ref{Fig:reheatn6}) top-right.

Interestingly if one increases the value of $y_\phi$ sufficiently, there also exists the possibility
that the inflaton decays {\it before} the PBHs population. In this case, 
the first phase of the reheating is dominated by the inflaton decay process.
This phase is achieved at a time $t\simeq \Gamma_\phi^{-1}$ 
with a Universe dominated by radiation. However, in a second phase, 
as $\rho_R\propto a^{-4}$ whereas $\rho_{\BH} \propto a^{-3}$, at a given time the PBH energy density will surpass the radiation density, driving the expansion rate. Finally, they will release their entropy through their decay in a third phase, all the radiation being then generated by the PBH. 
We illustrate this possibility in the lower left panels of Fig.(\ref{Fig:reheatn4}) and (\ref{Fig:reheatn6}), for $y_\phi=0.1$.
We clearly distinguish the 4 phases (inflaton-radiation-PBH-radiation),
where the inflaton decay for $a/\ae \simeq 100$, in accordance with Eq.(\ref{Eq:arhnobh}). Increasing $y_\phi$ further only reduces the inflaton
domination region as one can see in Fig.(\ref{Fig:reheatn4}) bottom-right.
\subsection{Inflaton reheating versus PBH reheating}
\begin{figure}[!]
\centering
\includegraphics[width=\columnwidth]{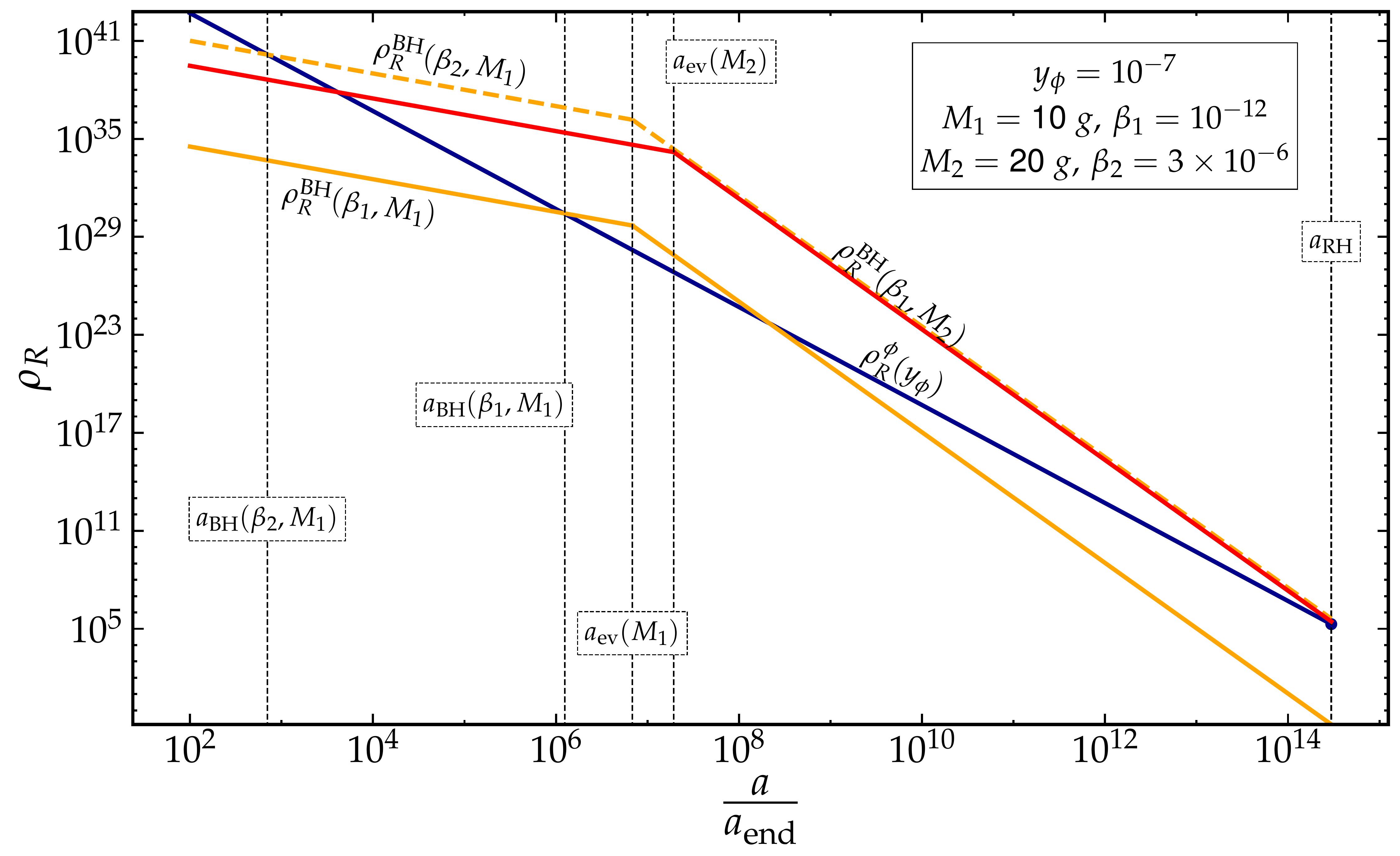}
\caption{\em Evolution of the density of radiation $\rho_R^{\BH}$ (orange and red)generated by different populations of PBHs ($\beta_i$,$\Min$) in comparison with the radiation produced by the inflaton decay, $\rho_R^\phi$ in blue for $V(\phi)\propto \phi^4$. 
We observe that increasing $\beta$
{\it or} $\Min$ allowed for PBH reheating domination.}
\label{Fig:pbhillustration}
\end{figure}

One can then compute, for each value of $y_\phi$ the
proportion $\beta$ and mass $\Min$ for which the PBHs population begins to dominate the reheating process.
We illustrate the possibilities on  Fig.(\ref{Fig:pbhillustration}) where we plotted 
the evolution of radiation density
$\rho_R^{\BH}$ 
generated by a population of PBH ($\beta_i$, $\Min=M_i$) 
in comparison with the radiation produced by the inflaton
decay, $\rho_R^\phi$, in the case of a quartic potential
with $y_\phi=10^{-7}$.
Looking carefully at the figure, we understand that, whereas
the original set of parameter ($\beta_1=10^{-12}$, $M_1=10$ g) is not sufficient for $\rho_R^{\BH}$
to dominate before the end of the solely inflaton driven reheating process at $\arh$, increasing $\beta$ up
to $\beta_2=3 \times 10^{-6}$ {\it or} the PBHs mass
$\Min$ to $M_2=20$ g is sufficient to increase 
the energy transferred (larger $\beta$) or delay the 
PBH decay (larger $\Min$) such that $\rho_R^{\BH}=\rho_R^{\phi}$ at $\arh$. These are the threshold values
for a PBH reheating.

We can then compute analytically the corresponding $\beta(\Min)$
for each $y_\phi$ where the domination of PBH over the inflaton 
in the reheating process occurs. Considering that the Universe
is still dominated by the inflaton\footnote{We checked that this is always the case.}, we should ask that the radiation 
produced from the PBH decay at $\arh$ is larger than the radiation
produced by the inflaton at $\arh$. While $\rho_R^{\BH}$ follows
Eq.(\ref{Eq:rhorfrombh}) from $\abh$ till $\aev$, after the 
PBH decay, $\rho_R^{\BH}$ redshifts as $a^{-4}$. The condition
for a PBH-driven reheating should then be written
\beq
\rho_R^{\BH}(\aev)\left(\frac{\aev}{\arh}\right)^4 >\rhorh \,,
\eeq
\noindent
with $\rhorh$ given by Eq.(\ref{Eq:arhnobh}). We obtained for $n<7$
\bea
\beta^{n<7} &\gtrsim& \beta_{\rm crit}^{\phi}=\delta \times \left(\frac{y^2_\phi}{8\pi}\right)^{\frac{6w_\phi-2}{3-3w_\phi}}\left(\frac{M_P}{\Min}\right)^{\frac{2-2w_\phi}{1+w_\phi}}\nonumber
\\
&&
\times\,\lambda^{\frac{3w_\phi-1}{3w_\phi+3}}\left(\frac{\alpha_n}{M_P^4}\right)^{\frac{6w_\phi-2}{3-3w_\phi}}\,,
\label{Eq:betacrit}
\eea
and for $n>7$, we have
\bea
\beta^{n>7} &\gtrsim & \beta_{\rm crit}^{\phi}=\delta \times  \left(\frac{-\alpha_n}{M_p^4} \right) \left(\frac{y^2_\phi}{8\pi}\right)\left(\frac{M_P}{\Min}\right)^{\frac{2-2w_\phi}{1+w_\phi}}\nonumber
\label{Eq:betacritng7}
\\
&&
\times \, \lambda^{\frac{1-w_\phi}{2w_\phi+2}}\,\left(\frac{\rho_{\rm end}}{M_P^4}\right)^{\frac{9w_\phi-5}{6+6w_\phi}}
\eea
with
\beq
\delta= \frac{5+3w_\phi}{2\sqrt{3}\epsilon}(4\pi\sqrt{3} \gamma)^{\frac{-2w_\phi}{1+w_\phi}} \left[\frac{2\sqrt{3} \epsilon}{1+w_\phi}\right]^{\frac{5+3w_\phi}{3+3w_\phi}}  .
\eeq

\begin{figure*}[t!]
	\includegraphics[height=4.1cm]{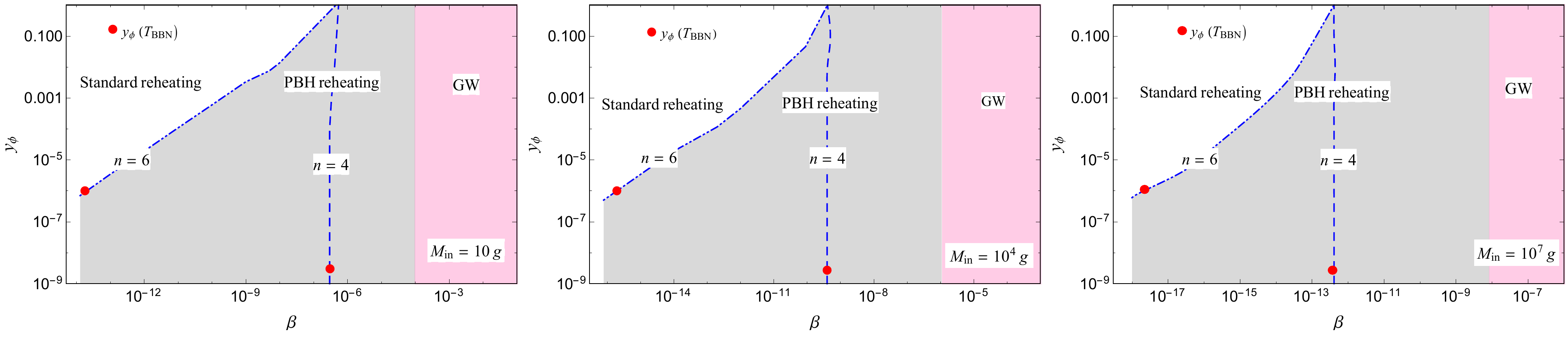}
\caption{\em Critical Yukawa coupling $y_\phi$ as function of $\beta$ for different $M_{\rm in}$ ($10$g, $10^4$g and $10^7$g, from left to right) and for $n=4$ (dashed) and $n=6$
(dot-dashed), corresponding to $w_\phi=\frac{1}{3}$ and $\frac{1}{2}$ respectively. Points in the shadow regions are subject to PBH reheating. Points in the extreme right (pink) region are excluded by an excess of gravitational wave, see Eq.(\ref{Eq:gw}). The red circle indicates the Yukawa coupling associated with the BBN temperature and the value of $y_\phi(T_{\rm BBN})=(1.8\times 10^{-9},\,7\times 10^{-7})$ for $n=(4,\,6)$ respectively.}
\label{Fig:yvsbeta}
\end{figure*}

\noindent
We note that for a quartic potential ($n=4$, $w_\phi={1}/{3}$),
the value of $\beta$ is independent of the Yukawa coupling $y_\phi$
and is $\beta \gtrsim 10^{-7}$, whereas for $n=6$
($w_\phi={1}/{2}$), the critical value of $\beta$ follows $\beta\propto y_\phi^{\frac{4}{3}}$.
We illustrate our result in Fig.(\ref{Fig:yvsbeta}) where we plotted
the minimal value of $y_\phi$ necessary for the 
inflaton to dominate the reheating process for a given $\beta$ in the case of PBH mass of
$\Min=10$g. The value of $y_\phi(\beta)$ below which the PBH reheating {\it have to} be taken into account to settle a coherent
thermal history of the early Universe is one of the main results of our work. In other words, for any given value of $\beta$, one
needs to check if the reheating process driven by the inflaton 
is not perturbed by the presence, and decay of the PBHs population
of mass $\Min$.

We also remark in Eq.(\ref{Eq:betacrit}) that the larger 
is the value of $\Min$, the smaller is be the value of $\beta_{\rm crit}^{\phi}$ necessary to realise the PBH reheating. Indeed, from
our discussion of Fig.(\ref{Fig:pbhillustration}) we understood that heavier PBHs have a tendency to decay later, injecting their (larger) entropy at a time when the inflaton is more diluted, facilitating in this way the domination of the PBH reheating over the inflaton reheating.
Their density of population, proportional to $\beta$, 
does not need to be so large then. Note that, for $n=6$ and $y_\phi=10^{-5}$, giving a reheating temperature of the order of $\sim 1$ GeV, already for
$\beta$ as small as $\sim 10^{-12}$, the PBH population dominates the reheating process. This 
important result should affects considerably, for instance,  gravitational reheating \cite{Haque:2022kez,Clery:2021bwz,Clery:2022wib}, which we defer for future study.

\subsection{PBH reheating temperature}

We have now all the tools in hand to compute the corresponding reheating  temperature in the different possible regimes ($\beta$,\,$\Min$) of PBHs population for a given $y_\phi$.
In one case PBHs, without being the dominant energy component, may populate the thermal bath through their decay, 
the reheating temperature is then given by the value of $\rho_R^{\BH}$ at $\arh$, when the PBHs evaporate. We obtain
\bea
&&
\rhorh=\rho_R^{\rm BH}(\aev)\times \left(\frac{\aev}{\arh}\right)^4 \nonumber \label{Eq:rhoRHsmallbetainflaton}\\
&&
\simeq \delta^{\frac{3(w_\phi+1)}{1-3\,w_\phi}}\,M_P^4 \left(\frac{M_P}{\Min}\right)^{\frac{6-6w_\phi}{1-3w_\phi}}
\beta^{\frac{3+3w_\phi}{3w_\phi-1}}\,.
\eea
From this, one can find
\\
\beq
\trh \simeq M_P \, \left[\frac{\delta^{\frac{3(w_\phi+1)}{4(1-3\,w_\phi)}}}{{\alpha_T}^{\frac{1}{4}}}\right] 
\left(\frac{M_P}{\Min}\right)^{\frac{3(1-w_\phi)}{2(1-3w_\phi)}}\beta^{\frac{3+3w_\phi}{12w_\phi-4}}\,,
\label{Eq:trhintermediate}
 \eeq

\noindent
where we have taken into account the redshift regime 
$\rho_R\propto a^{-4}$ after the PBH evaporation.

\begin{figure*}[!ht]
	\includegraphics[height=6.25cm]{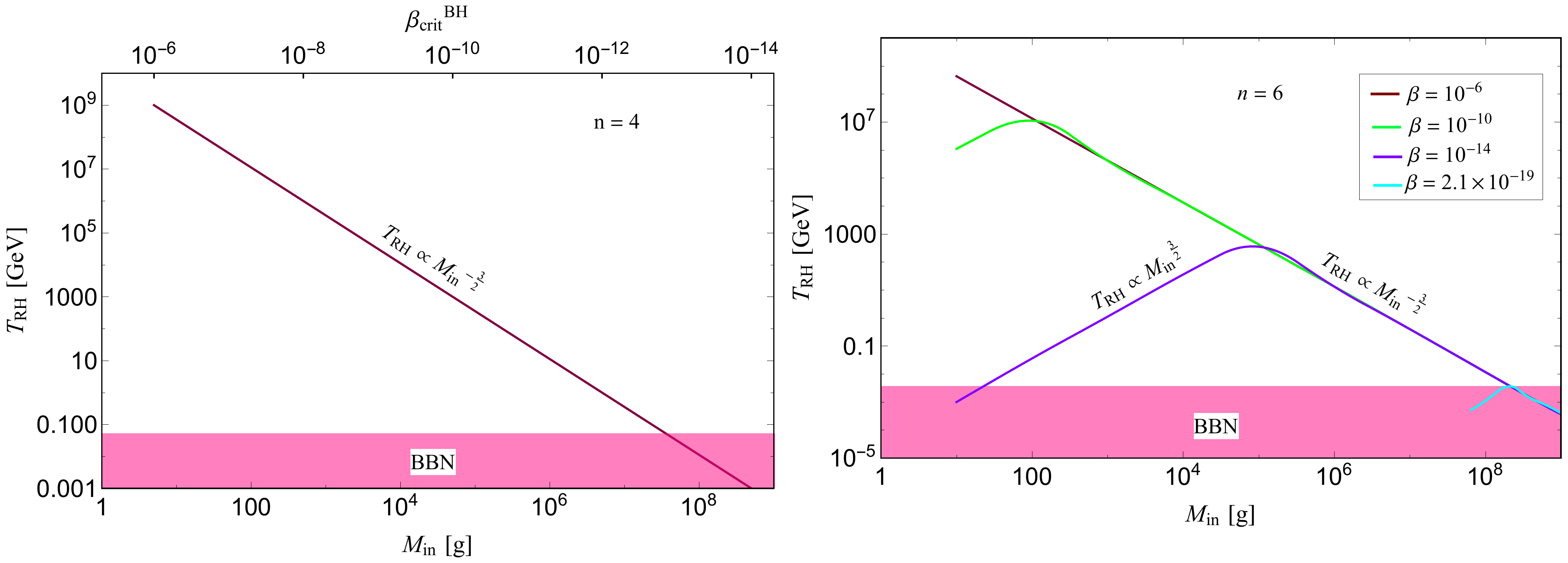}
	\caption{\em Reheating temperature as a function of initial BH mass $M_{\rm in}$ for two different values of $n=(4,\,6)$. In the left panel ($n=4$), the reheating temperature is independent of $\beta$, and reheating through BH evaporation is only possible when $\beta>\beta_{\rm crit}^{BH}$, which is shown in the upper panel of Fig. In the right panel, results are for $n=6$ with different values of $\beta=(10^{-6},\,10^{-10},\,10^{-14},\,2.1\times 10^{-19})$. 
 Points in the shaded (pink) region are excluded by BBN bounds, Eq.(\ref{upper-bound-beta}).} 
	\label{Fig:PBHtre}
\end{figure*}

In the second case, $\beta$ crossing certain threshold value, Eq.(\ref{Eq:betamin}), 
permits the PBH to dominate the Universe's energy budget over the inflaton field. Finally, reheating ends with PBHs decay again at $\aev$, 
and all the entropy generated by their decay are transferred to the thermal bath. This happens for
\beq
\Gamma_{\rm BH} = H ~~\Rightarrow \rhorh=3 M_P^2\Gamma_{\rm BH}^2\,,
\eeq
or, with $\Gamma_{\rm BH}= {\epsilon} \frac{M_P^4}{\Min^3}$
\beq
\rhorh=3\,{\epsilon^2}\frac{M_P^{10}}{\Min^6}\,,
\label{Eq:rhoRHlargebeta}
\eeq
which gives
\beq
\trh= M_P\,\left(\frac{3\,{\epsilon}^2}{\alpha_T}\right)^{\frac{1}{4}} \left(\frac{M_P}{\Min}\right)^{\frac{3}{2}}\,.
\label{Eq:largebeta}
\eeq

This is another important result of our paper.
We illustrate it in Fig.(\ref{Fig:PBHtre}) where we show the
reheating temperature $\trh$ as a function of $\Min$ in the case $n=4$ (left) and $n=6$ (right) after solving numerically the set of Boltzmann equations. 
In the scenario of $\beta>\beta_{\rm crit}^{\rm BH}$, PBH reheating happens after PBH domination, and the dependence on $\beta$ and $w_\phi$ disappears as we can see from Eq.(\ref{Eq:largebeta}). Here,
we need a minimum value $\abh$ given by Eq.(\ref{Eq:abh})
to reach the regime of PBH domination and follow the dependency
$\trh \propto \Min^{-\frac{3}{2}}$ which is effectively what we observe in both the left and right panel of Fig.(\ref{Fig:PBHtre}). 
However, if PBHs evaporate during inflation domination, we have a $w_\phi$ dependent behavior in $T_{\rm RH} (M_{\rm in})$ given by Eq.(\ref{Eq:trhintermediate}), $T_{\rm RH}\propto M_{\rm in}^{3(1-w_\phi)/2(3w_\phi-1)}$. As an example for $w_\phi=1/2$, $T_{\rm RH}\propto M_{\rm in}^{3/2}$. 
It should be noted that in this scenario, there is a threshold value for $\beta$, below which the PBH reheating scenario cannot achieve a reheating temperature higher than the energy scale of BBN, which is about 5 MeV.
This threshold value can be calculated from Eq.(\ref{Eq:betamin}) upon plugging $M_{\rm in} = M_{\rm in}^{\rm max}\sim 2\times 10^{8}$ g (see Eq.(\ref{minmax}) of appendix \ref{appendixB}).
For $w_\phi=1/2$, this value turns out to be $2.1 \times 10^{-19}$.

\begin{figure*}[!ht] 
\includegraphics[height=19.6cm]{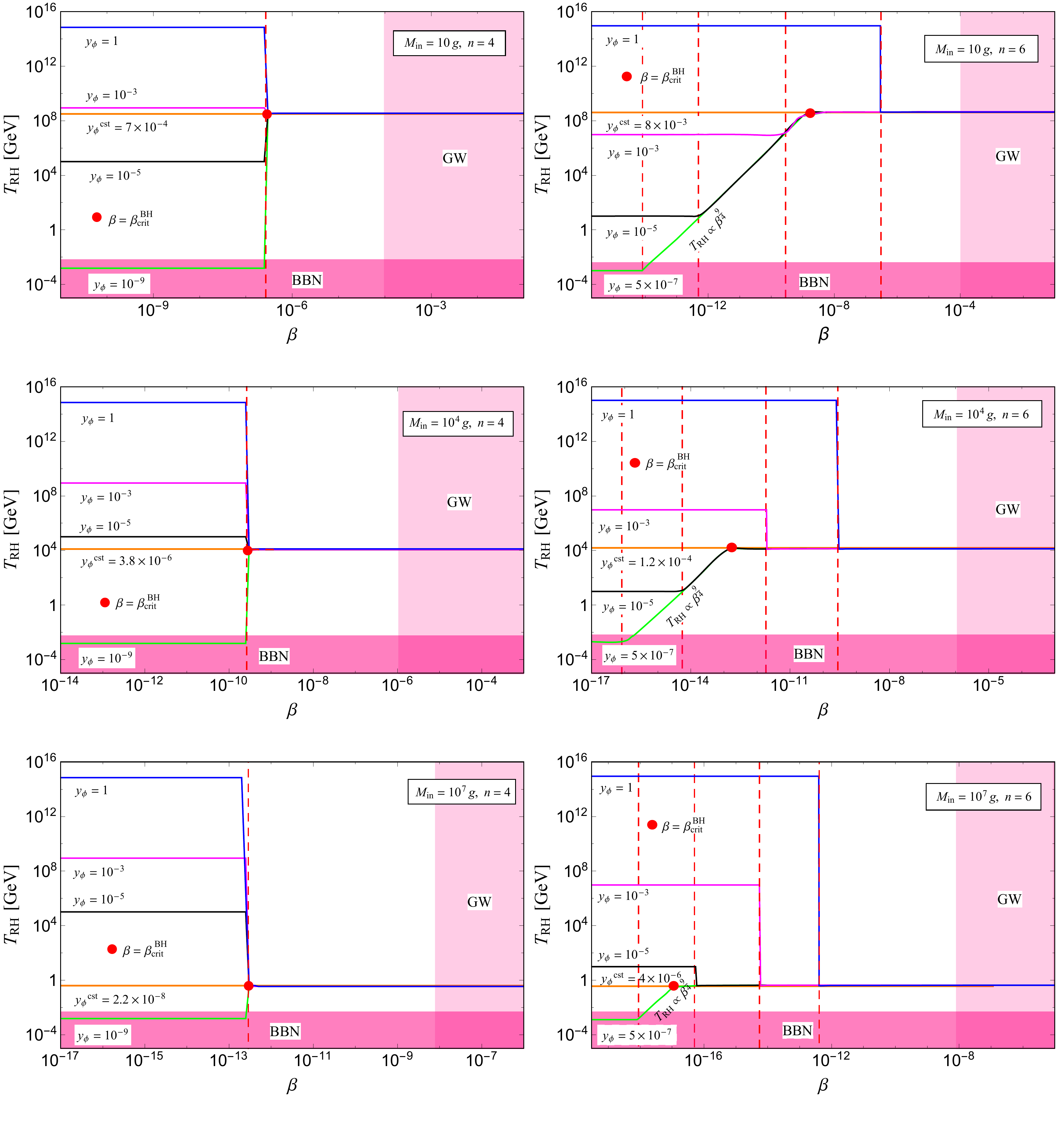}
	\caption{\em Evolution of the reheating temperature $\trh$ as function of $\beta$ for different values of $y_\phi$ with $\Min=10, 10^4, 10^7$g (from top-down)
 and $n=4$ and 6 (left column and right column). The solid red circle indicates the critical $\beta =\beta_{\rm crit}^{\rm BH}$ defined in Eq.(\ref{Eq:betamin}). Vertical red dashed lines on the left of the solid red circle indicate $\beta =\beta_{\rm crit}^{\phi}$, whereas those on the right indicate critical $\beta$ values at which the Universe undergoes a transition from inflaton $\to$ radiation $\to$ PBHs domination.
The extreme right light-pink shaded regions are excluded by an excess of gravitational wave, Eq.(\ref{Eq:gw}), whereas dark-pink shaded lower regions are excluded by BBN bounds, Eq.(\ref{upper-bound-beta}).}


	\label{Fig:reheatn6final}
\end{figure*}

Finally, we summarize all our analyses in the set of Figs. (\ref{Fig:reheatn6final}) where we plotted the reheating temperatures obtained as a function of $\beta$ for different values of $\Min$ and $y_\phi$ in the cases of $n=4$ and $n=6$.
To facilitate our discussion further, we identify a critical coupling, $y^{\rm cst}_{\phi}$, defined by equating radiation energy density at the end of PBH domination, Eq.(\ref{Eq:rhoRHlargebeta}), with that at the end of standard inflaton domination, Eq.(\ref{Eq:arhnobh}). This value $y_\phi^{\rm cst}$ can be considered as the coupling 
needed for the inflaton to reheat as efficiently as 
a PBHs population of mass $\Min$, and is given by
\beq
y^{\rm cst}_{\phi}|_{n<7} = \sqrt{8\pi} \left(\frac{M_P^4}{\alpha_n} \right)^{\frac{1}{2}} \left(\frac{3\epsilon^2}{\lambda} \right)^{\frac{1-w_\phi}{4(1+w_\phi)}} \left(\frac{M_P}{M_{\rm in}} \right)^{\frac{3(1-w_\phi)}{2(1+w_\phi)}}\,,
\label{Eq:yukconstantTRH}
\eeq
 The above expression is deduced for $n<7$. On the other hand, the critical Yukawa coupling for $n>7$ would be
 
\bea
y^{\rm cst}_{\phi}|_{n>7}&=& \sqrt{8\pi} \left(\frac{M_P^4}{-\alpha_n} \right)^{\frac{1}{2}} \lambda^{\frac{w_\phi-1}{4+4w_\phi}} \left(\frac{ 3\epsilon^2M_P^6}{M_{\rm in}^6} \right)^{\frac{3w_\phi-1}{6+6w_\phi}}
\nonumber\\
&& \times \, \left(\frac{\rho_{\rm end}}{M_P^4}\right)^{\frac{5-9w_\phi}{12+12w_\phi}} .
\label{Eq:yukconstantTRHng7}
\eea


 For example, $n=4$ ($6$) and $M_{\rm in} = 10$ g, one can deduce $y^{\rm cst}_{\phi} = 7\times 10^{-4}$ 
($8\times 10^{-3}$), while for $M_{\rm in} = 10^4$ g, 
the critical coupling reduces to $3.8\times 10^{-6}$ ($1.2\times 10^{-4}$). 
Associated with those two mass values, reheating temperatures are deduced from Eq.(\ref{Eq:largebeta}), 
$ T_{\rm RH}^{\rm cst} = (4.2\times 10^{8}, 1.3\times 10^{4})$ GeV for $M_{\rm in}=(10, 10^4)$ g respectively. 
It is interesting that those temperatures are independent 
of all inflationary parameters such as $(n,y_\phi)$ as expected. These features are recovered from full numerical analysis and clearly depicted in Fig.(\ref{Fig:reheatn6final}), where $y_\phi = y^{\rm cst}_{\phi}$ corresponds to constant say reheating temperature $T_{\rm RH} = T_{\rm RH}^{\rm cst}$ line along which all the different curves meet behaving like an attractor which erases all the microscopic information about the inflaton. The significance of such behavior could be interesting to look into.   


For a given $y_\phi \leq y^{\rm cst}_{\phi}$ and $n>4$, we observe in the right panels of Fig.(\ref{Fig:reheatn6final})
three distinct regions along $\beta$ with two constant temperature plateaus and one with constant slope. 
For smaller $\beta < \beta_{\rm crit}^{\phi}$ in the first plateau region, reheating is solely inflaton driven with constant reheating temperature for a fixed coupling $y_\phi$ respecting the Eq.(\ref{Eq:trhwobh}). 
In the intermediate regime of $\beta_{\rm crit}^{\phi} < \beta < \beta_{\rm crit}^{\rm BH}$, the radiation contribution from PBHs evaporation takes over the inflaton radiation, and hence the reheating temperature varies with a constant slope 
$\trh\propto \beta^{\frac{3+3w_\phi}{12w_\phi-4}} =\beta^{9/4}$ for $w_\phi=\frac{1}{2}$, as expected from Eq.(\ref{Eq:trhintermediate}). 
Such a slope is observed in the green line 
for all three PBHs initial mass values,
black line for $M_{\rm in} = (10, 10^4)$ g, and magenta line for $M_{\rm in} = 10$ g shown in the right panel of Fig.(\ref{Fig:reheatn6final}).

Finally, for a large value of $\beta > \beta_{\rm crit}^{\rm BH}$, the PBHs itself dominates over the inflaton, and subsequent decay leads to reheating temperature hitting the $T_{\rm RH} = T_{\rm RH}^{\rm cst}$ attractor line respecting Eq.(\ref{Eq:largebeta}). The reheating temperatures are deduced to be $T_{\rm RH} \simeq (4.2\times 10^{8}, 1.3\times 10^{4}, 0.4)$ GeV for $M_{\rm in} = (10, 10^4, 10^7)$ g, respectively that can be recovered from the plots.

However, for a given $y_\phi \leq y^{\rm cst}_{\phi}$, n = 4 ($w_\phi=1/3$) deserves special attention, 
because intermediate regime $\beta_{\rm crit}^{\phi} < \beta < \beta_{\rm crit}^{\rm BH}$ does not exists. The reason behind this feature is that for $n=4$, both inflaton and radiation energy densities dilute in a similar manner ($\propto a^{-4}$), and that leads to no PBH reheating for those PBHs evaporate during inflaton domination.  The feature is similar to the case $y_\phi > y^{\rm cst}_{\phi}$ discussed below; the only difference is that for $y_\phi>y_{\phi}^{\rm cst}$, before PBH domination the universe always remained radiation dominated.

 Once, $y_\phi > y^{\rm cst}_\phi$, Fig.(\ref{Fig:reheatn6final}) shows two plateau regions of reheating temperature with an abrupt fall at a new $\beta$ critical value depicted again by the vertical red dashed lines placed at the right side of the $\beta_{\rm crit}^{\rm BH}$ red circle. The first plateau indicates the fact that due to strong inflaton coupling, below this new critical $\beta$ value reheating is governed solely by inflaton without any significant effect from PBHs, leading to $\beta$-independent $T_{\rm RH}$, followed by Eq.(\ref{Eq:trhwobh}). However, once one assumes the new critical value of $\beta$, or higher, the universe undergoes  from inflaton $\to$ radiation $\to$ PBH, and then after PBHs evaporation leads to again $(\beta,y_\phi)$ independent reheating temperature hitting the $T_{\rm RH} = T_{\rm RH}^{\rm cst}$ attractor line.

Finally, it has been recently observed that Hawking evaporation during PBH domination leads to a small-scale cosmological fluctuation that, in turn, provides an induced stochastic gravitational wave background. This GW background could provide a stronger constrain on the $\beta$ parameter coming from BBN \cite{Domenech:2020ssp} :

\beq \beta< 1.1\times 10^{-6}\left(\frac{w^{3/2}}{0.2}\right)^{-\frac{1}{2}}\left(\frac{M_{\rm in}}{10^{4}\, {\rm g}} \right)^{-17/24}\,.
\label{Eq:gw}
\eeq
As an example   for $M_{\rm in}=10$ g, to satisfied this constrain, $\beta$ must be $< 10^{4}$. We added these constraints in Figs.(\ref{Fig:yvsbeta}) and (\ref{Fig:reheatn6final}).

\section{The case for extended mass distribution}


The extended mass function (EMF) of PBHs is intricately tied to the underlying mechanism that governs their formation, and are contingent on the power spectrum of primordial density perturbations and the equation of state of the Universe at the time of their formation (see Ref.~\cite{Carr:2017jsz}). Consequently, distinct shapes of the mass function $f_{\rm PBH}(M,t)$ emerge: power-law~\cite{Carr:1975qj}, log-normal \cite{Dolgov:1992pu,Green:2016xgy,Dolgov:2008wu}, critical collapse \cite{Carr:2016hva,Musco:2012au,Niemeyer:1999ak,Yokoyama:1998xd}, or metric preheating \cite{Martin:2019nuw,Martin:2020fgl,Auclair:2020csm}, among others.

In this work, we consider the class of PBHs with power-law shape mass function. This type of mass function corresponds to the scenario where the PBHs form from scale-invariant fluctuations, that is, with constant amplitude at the horizon epoch. This happens when the Universe is dominated by a perfect fluid with the constant equation of state. The concerned mass function at the initial time, $t_i$ is given by:
\beqa
f_{\rm PBH}(M_i,t_i) = \begin{cases}
    C M_i^{-\alpha}, & {\rm for}\, \,  M_{\rm min} \leq M_i \leq M_{\rm max} \, \\
    0, & {\rm otherwise}\, .
\end{cases}
\label{Eq:power-law-dist}
\eeqa
The coefficient $C$ is the overall normalization factor, and $(M_{\rm min}$, $M_{\rm max})$ represents the minimum and the maximum PBH masses, respectively. They depend on the domain of frequency over which the scale-invariant fluctuations are formed. Subsequently, we assume that the distribution extends to lower masses, hence we set $M_{\rm max} = M_{\rm in}$, where $M_{\rm in}$ is given in Eq.~(\ref{Eq:pbh-m-in}),  and treat $M_{\rm min}$ as a free parameter. The parameter $\alpha$ depends on the equation of state at formation, $P=w \rho$, and is given by \cite{Cheek:2022mmy}:
\beqa
\alpha = \frac{2+4w}{1+w}\, .
\eeqa

Concerning the evolution of energy densities, recall that like in the monochromatic case, our analysis pertains to a dynamical system that consists of an oscillating inflaton field, whose evolution is described by the Eq.~(\ref{Eq:rhophi}), and evaporating PBHs. 
Thus, before the complete evaporation the PBHs, the two sources of the background radiation are the inflaton and the PBHs. Note however that PBHs leaving behind remnant masses is a possibility, as argued in Ref.~\cite{Dalianis:2021dbs}.

The comprehensive treatment of the PBH mass and spin distributions, as well as their cosmological evolution, can be found in Refs.\cite{Dienes:2022zgd,Cheek:2022mmy}, where the relevant evolution equations have been derived. Nonetheless, for consistency, we summarize the main equations used in the current work. The number and energy density of PBHs, at time $t$ can be written respectively as:
\beqa 
n_{\rm BH}(t) &=& \int_{0}^{\infty} f_{\rm PBH}(M, t) dM \, ,
\label{Eq:pbh-number-density} \\
\rho_{\rm BH}(t) &=& \int_{0}^{\infty} M f_{\rm PBH}(M, t) dM \,.
\label{Eq:pbh-energy-density}
\eeqa  
The mass spectra $f_{\rm PBH}(M, t)$ at time $t$ can be related to $f_{\rm PBH}(M_i,t_i)$. Indeed, upon establishing an initial spectrum $f_{\rm PBH}(M_i, t_i)$ at time $t_i$, the distribution undergoes changes due to both cosmic expansion and evaporation. Nonetheless, the comoving number density of PBHs with initial masses within an infinitesimal range of [$M_i, M_i + dM_i$] remains constant until the time when they completely evaporate, resulting in a drop of the number density to zero. One can then express $f_{\rm PBH}(M, t)$ as follows:
\beqa 
a^3 f_{\rm PBH}(M, t) dM = a^3_{\rm in} f_{\rm PBH}(M_i, t_i) dM_i
\label{Eq:conservation_conum_density}
\eeqa  

It follows from Eq.(\ref{Eq:pbh-energy-density}), that the Friedmann-Boltzmann equation for $\rho_{\rm BH}(t)$ is given by
\beqa 
 \dot{\rho}_{\rm BH} + 3 H \rho_{\rm BH} = \frac{a^3_{\rm in}}{a^3} \int_{\widetilde M}^{\infty} \frac{dM}{dt} f_{\rm PBH}(M_i, t_i) dM_i \, ,
\label{Eq:evolution-eq-pbh}
\eeqa 
where $dM/dt$ describes the rate of change of PBH mass $M$ due to evaporation, and the lower bound $\widetilde M$ allows to ensure that at time $t$ only the non-evaporated PBHs with mass $M_i > \widetilde M (t)$ contribute to their energy density. Using Eq.(\ref{Eq:mbh}), $\widetilde M (a)$ can be estimated as
\beqa 
\widetilde M(a) = \left(\frac{2\sqrt{3} \epsilon}{1+w_\phi} \right)^{1/3} \left(\frac{M_P^5}{\sqrt{\rho_{\rm end}}} \right)^{1/3} \left(\frac {a}{a_{\rm end}}\right)^{\frac{1}{2}(1+w_\phi)}.
\label{eq:eva_ini_mass}
\eeqa 
Hence, it is then straightforward to derive the evolution of the radiation as:
\begin{multline}
\dot{\rho}_R + 4H\rho_R = \Gamma_{\phi}\rho_{\phi}(1+w_\phi) \\ -  \frac{a^3_{\rm in}}{a^3} \int_{\widetilde M}^{\infty} \frac{dM}{dt} f_{\rm PBH}(M_i, t_i) dM_i 
\label{Eq:evolution-eq-rad}
\end{multline}

In sum the evolving system can be described by Eq.(\ref{Eq:rhophi}),  (\ref{Eq:evolution-eq-pbh}), and (\ref{Eq:evolution-eq-rad}), together with Eq.(\ref{Eq:hub}). Solving this set of equations is somehow subtle compared with the monochromatic case. The integrals in the right-hand sides of Eqs.(\ref{Eq:evolution-eq-pbh}) and (\ref{Eq:evolution-eq-rad}) must be evaluated at every time $t$, which requires a different approach. Thus, regarding the methodology, we utilized a modified version of the package \hyperlink{https://github.com/yfperezg/frisbhee}{\tt FRISBHEE}\footnote{\hyperlink{https://github.com/yfperezg/frisbhee}{https://github.com/yfperezg/frisbhee}} \cite{Cheek:2021odj,Cheek:2021cfe,Cheek:2022dbx,Cheek:2022mmy}, to include inflaton to the evolving system. Further details on the numerical approach used to solve these equations can be found in \cite{Cheek:2022mmy}.

\begin{figure}[!ht]
 \centering
	\includegraphics[width=1.05\columnwidth]{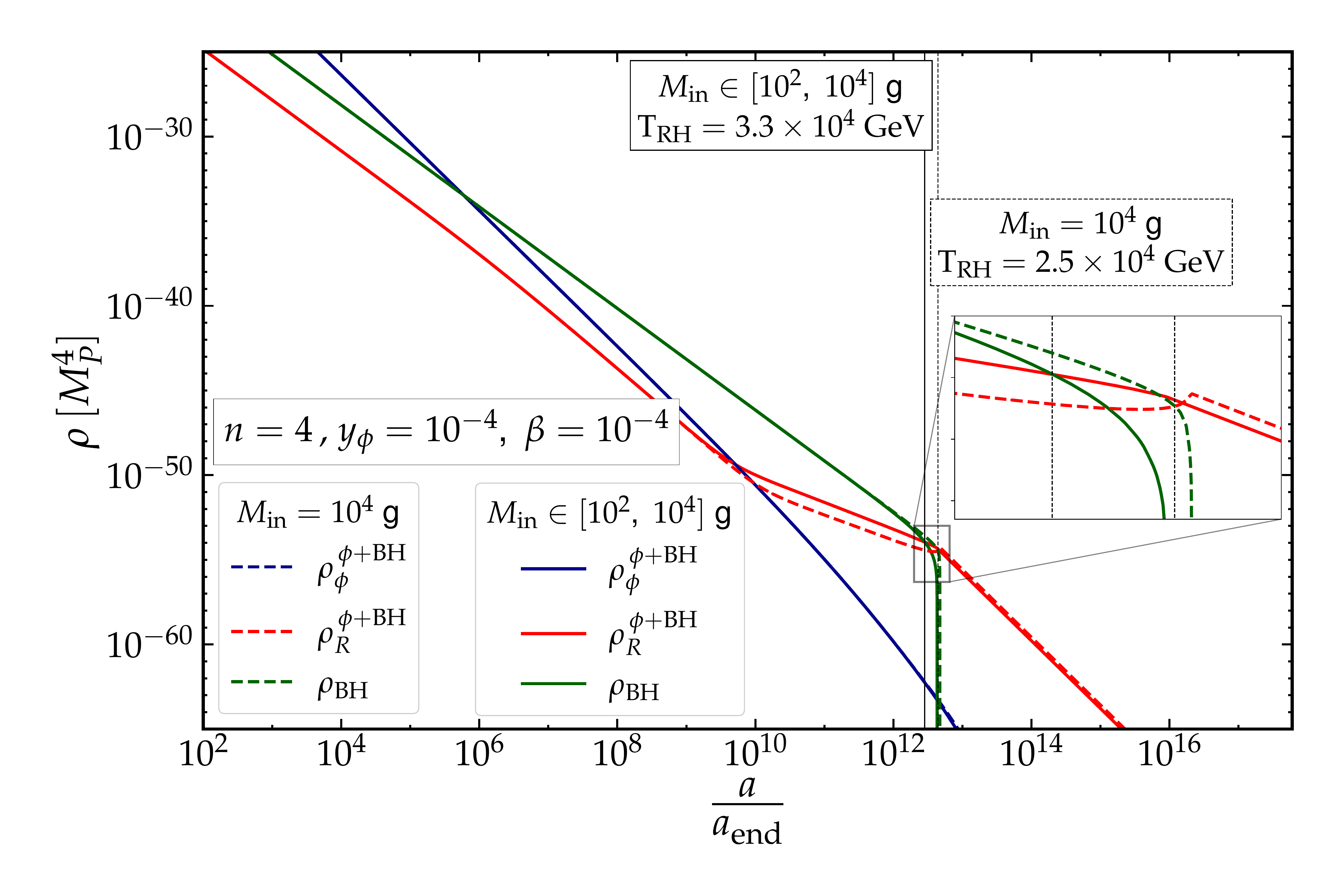}
	\caption{Evolution of the energy densities $\rho_{\phi}$ (blue), $\rho_{R}$ (red) and $\rho_{\rm BH}$ (green) as function of $a/a_{\rm end}$ for $n=4, y_\phi=10^{-4}, \beta=10^{-4}$, for monochromatic limit $M_{\rm in}\approx 10^{4}$ g (dashed), and extended distribution, $M_{\rm in} \in $ [$10^{2}, 10^{4}$] g (full). The reheating temperatures in both cases are very close to each other.} 
	\label{Fig_k=4-M1E4_Mono_pbh_ext_vs_mono}
\end{figure}

With regards to the findings pertaining to the EMF case, a comparable reheating temperature is obtained as that of monochromatic PBHs mass spectra, as exemplified in 
Fig.(\ref{Fig_k=4-M1E4_Mono_pbh_ext_vs_mono}), for $n=4, y_\phi=10^{-4}$, and $\beta=10^{-4}$. 
This plot corresponds to the monochromatic limit with mass $M_{\rm in}\approx 10^{4}$ g, and for the extended distribution with $M_{\rm in} \in $ [$10^{2}, 10^{4}$] g. Note that typically, the reheating through PBHs, after a regime of PBH domination, happens when they completely evaporate. Hence, in our case, since we have chosen the mass function that extends to lower values, with the maximal initial mass $M_{\rm max}$ corresponding to the monochromatic mass $M_{\rm in}$, we expect the complete evaporation of PBHs in both cases is achieved at the same epoch.

\noindent
Nevertheless, it is worth pointing out that the reheating occurs slightly earlier in the case of EMF  as can be seen in Fig.(\ref{Fig_k=4-M1E4_Mono_pbh_ext_vs_mono}), for $a/a_{\rm end}$ in about $10^{9}$ to the evaporation point. As soon as the lighter population of PBHs starts to evaporate and inject energy in the radiation, the total energy density of the radiation bath in the EMF case (solid red) would start becoming larger as compared to the monochromatic one (dashed red). Eventually, the former (EMF case) would lead to the reheating point that happens shortly before $\widetilde M$, defined in Eq.(\ref{eq:eva_ini_mass}), reaches $M_{\rm max}$, where, in turn, the reheating in the monochromatic case occurs.

{ 
\noindent
Also, it would be worth commenting that if the EMF extending to larger masses such that $M_{\rm in} = M_{\rm min}$, was considered, then the reheating temperature can be very affected depending on the width of the mass function, since the larger masses than $M_{\rm in}$ would have a longer lifetime compared to the monochromatic scenario.


We want to comment also that although in our study we focused on situation where the PBHs are formed during the inflaton domination phase, the alternative situations in
which they form during radiation domination era 
is another possibility depending on when they form ($M_{\rm in}$) and
$y_{\phi}$. In fact this situation arises when the formation of PBHs happens after the point where the standard reheating
is achieved, such that $M_{\rm in} > M^{\rm RH}_{\rm in}$, where $M^{\rm RH}_{\rm in}$ is:
\beqa
  M_{\rm in} > M^{\rm RH}_{\rm in} = 4\pi\gamma \sqrt{\frac{3}{\lambda}}  \left(\frac{8\pi}{y_{\phi}^2} \right)^{\frac{n}{2}} \left(\frac{M_P^4}{\alpha_n} \right)^{\frac{n}{2}} M_P
\label{Eq:MassBHformRadDom}
\eeqa
The case of $n=4$ and $y_{\phi}=2$, corresponding to $M^{\rm RH}_{\rm in}=1.6$ g, is illustrated by the lower right plot of Fig.(\ref{Fig:reheatn4}) where $M_{\rm in} \simeq 10$ g. For example, for $n=6$, $y_{\phi}=0.5$ ($0.1$), would lead to $M^{\rm RH}_{\rm in}\simeq 31$ g ($4.8\times 10^{5}$ g).
}

\section{Conclusion}

The proposal of forming PBHs in early universe cosmology has been the subject of intense investigations over the last decade. Most of the PBH studies so far are mainly either concentrated on the possible formation mechanism along with their late time effects, which the cosmological observation can constrain. In this paper, we explore the effect of PBHs in a mass range on which detailed exploration is still lacking, and furthermore, direct detection is less effective. In this paper, we particularly concentrate on the reheating phase after the inflation and propose a new mechanism of reheating, taking into consideration the effect of PBHs. 

As a case study, we consider the following main ingredients in our analysis. We consider the production of thermal baths from two production channels: inflation decay and PBH evaporation. Inflaton is creating the thermal bath through a Yukawa-type coupling with the Fermions, $y_\phi \phi \bar f f$, whereas
PBHs are assumed to be formed during the process of reheating parametrizing by their formation mass ($M_{\rm in}$) and initial abundance ($\beta$). Those PBHs can also evaporate and populate the thermal bath. Depending upon the value of those parameters, PBHs can potentially impact the reheating process. 

Considering a generic inflaton potential of the form $V(\phi)\propto \phi^n$, we analyzed the impact of the parameters ($\beta, \,\Min, \,y_{\phi}$) on the process of reheating in their respective regime. We show that the phenomenology of such scenarios is extremely rich, and depending on the values of those parameters, PBH can either dominate the reheating process or even change the expansion rate. We mostly focused on the impact of monochromatic PBH mass function. The inclusion of extended mass function did not produce any new features except minor quantitative changes in the physical quantities, such as reheating temperature.  

We discovered two distinct classes of reheating processes in addition to the conventional one driven purely by inflaton. If the inflaton coupling $y_\phi$ is lower than some critical value $y_\phi^{\rm cst}$ given by 
Eq.(\ref{Eq:yukconstantTRH}) in the limit of  $\beta\geq\beta_{\rm crit}^{\phi}$  for any $n\geq 4$ values, 
the radiation bath happened to be controlled by the evaporation of PBHs with its final temperature following the 
power law relation $T_{\rm RH} \propto \Min^{-\frac{3(1-w_\phi)}{2(1-3w_\phi)}}\beta^{\frac{3+3w_\phi}{12w_\phi-4}}$. This power law appears for a specific range of 
$\beta$ within  $\beta_{\rm crit}^{\phi} \leq \beta < \beta_{\rm crit}^{\rm BH}$. However, if the abundance 
assumes higher than $\beta_{\rm crit}^{\rm BH}$ and $n>2$, due to the slower rate of dilution of PBHs $\propto a^{-3}$ at certain time $a_{\rm BH}$, PBH dominates over the 
inflaton and leads to a universal reheating temperature $T_{\rm BH} \propto {\Min}^{-\frac{3}{2}}$ irrespective of $(w_\phi,\,\beta)$ values.
Interestingly, such a scenario behaves like an attractor in $(T_{\rm RH},\, \beta)$ plane, which erases all the initial information about inflaton and PBHs abundance. From the observational perspective, it would be interesting to look into such reheating scenario in greater detail.


Consequently, the reheating temperature can change drastically in the presence of PBHs. The real reason is that, whereas the inflaton dilutes faster than $\propto a^{-3}$ for $n>2$, the density of PBH still follows a dust-like evolution. Moreover, the radiation generated 
by the PBHs is also much less redshifted than the radiation produced by inflaton decay, opening the possibility of reheating driven by PBH
even if their density does not dominate over the inflaton. Our results are summarized in Fig.(\ref{Fig:reheatn6final}), where the reheating temperature in the presence of the primordial black hole is explicitly shown as a function of $\beta$ for different values of 
$y_\phi$ and $\Min$. 

\vspace{0.5cm} 
\noindent
\acknowledgements
E.K. and Y.M. want to thank L. Heurtier for extremely valuable discussions during the completion of our work. This project has received support from the European Union's Horizon 2020 research and innovation programme under the Marie Sklodowska-Curie grant agreement No 860881-HIDDeN, and the IN2P3 Master Projet UCMN. 
M.R.H wishes to acknowledge support from the Science and Engineering Research Board (SERB), Government
of India (GoI), for the SERB National Post-Doctoral fellowship, File Number: PDF/2022/002988. DM wishes
to acknowledge support from the Science and Engineering Research Board (SERB), Department of Science and
Technology (DST), Government of India (GoI), through the Core Research Grant CRG/2020/003664.  DM also thanks the Gravity and High Energy
Physics groups at IIT Guwahati for illuminating discussions.
The work of E.K. was supported by the grant "Margarita Salas" for the training of young doctors (CA1/RSUE/2021-00899), co-financed by the Ministry of Universities, the Recovery, Transformation and Resilience Plan, and the Autonomous University of Madrid.

\appendix

\section{The inflationary parameters}\label{appendixA}

An important feature of inflationary models is the possibility of reheating the Universe after the inflation, leading a radiation dominated epoch.
Inflation reheating refers to the process by which the energy of the inflaton field, which powered the inflationary expansion of the Universe, is transferred to other particles in the Universe. This transfer of energy occurs at the end of the inflationary period and is considered to have created the conditions necessary for the formation of structure in the Universe. 
The transfer of energy from the inflaton to other particles is thought to have been accomplished through a variety of mechanisms, such as the decay of the inflaton into other particles or the production of particles through the interaction of the inflaton with other fields~\cite{Abbott:1982hn,Dolgov:1982th,Nanopoulos:1983up,Garcia:2020wiy}. 

In the following study we assume that the reheating is not instantaneous, that is, a scenario in which the transfer of energy from the inflaton field to other particles at the end of inflation occurs over a longer period of time, rather than instantaneously. Note that there has been many works which have taken into account non-instantaneous reheating scenario (see for example Refs.~ \cite{Giudice:2000ex,Garcia:2017tuj,Dudas:2017rpa,Chen:2017kvz,Garcia:2020eof,Bernal:2020gzm,Co:2020xaf,Garcia:2020wiy}.

We start our discussion by considering a specific inflationary model, the so-called $\alpha-$attractor model, that permits a slow-roll inflation. The potential $V(\phi)$ has the following form:
    \begin{equation} \label{a}
V(\phi) = \Lambda^4 \left[  1 - e^{ -\alpha_1\dfrac{\phi}{M_P} } \right]^{n}\,,
\end{equation}
where $\alpha_1=\sqrt{\frac{2}{3\alpha}}$. The CMB power spectrum naturally fixes the mass scale $\Lambda$. Further, throughout our analysis, we consider $\alpha=1$. If we expand the above potential around minima, it can be expressed in a power law form as:
\be\label{potentialminima}
V(\phi) = \Lambda^4\left(\frac{\alpha_1}{M_P}\right)^{n}\,\phi^{n}=\lambda\frac{\phi^n}{M_P^{n-4}} \, ,
\ee
where $\lambda=\left(\frac{\Lambda}{M_P}\right)^4\,\alpha_1^n$. Using the constraints from the CMB, the parameter $\Lambda$ can be expressed  in terms of the CMB observables such as $A _{\mathcal{R}}$, $n _s$, and $r$ as (see,
for instance, Ref.\cite{Drewes:2017fmn})
\beqa
 \lambda = & &\alpha_1^n {\left(\frac{3\pi^2 r A_{\mathcal{R}}}{2}\right)^4 }  \nonumber \\
 & & \times \left[\frac{n^2+n+\sqrt{n^2+3\alpha(2+n)(1-n_s)}}{n(2+n)}\right]^{n}
 \eeqa
 $A_{\mathcal{R}}\sim 2.19\times10^{-9}$ represents the amplitude of the inflaton fluctuation measured from Planck \cite{Planck:2018jri}. 
From the condition on the end of the inflation,
\beqa
\epsilon_v(\phi_{\rm end})=\frac{1}{2\, M_P^2}\left(\frac{V'(\phi)}{V(\phi)}\Big|_{\phi=\phi_{\rm end}}\right)^2=1\, ,  \nonumber
\eeqa
the field value at the end of the inflation can be written as
\bea\label{infpotentialfield}
\phi_{\rm end}=\frac{M_P}{\alpha_1}~ \ln\left(\frac{n}{\sqrt{3\alpha}}+1\right)\,.
\eea
Upon substitution of the above Eq.(\ref{infpotentialfield}) into Eq.(\ref{potentialminima}), the expression of the potential at the end of inflation takes the following form
\be
V(\phi_{\rm end})=\frac{\lambda\,M_P^4}{\alpha_1^4} \left(\frac{n}{n+\sqrt{3\alpha}}\right)^{n} \, .
\ee
Finally, the inflaton energy density at the end of inflation, which provides the initial condition for the subsequent reheating dynamics, turns out as (using the condition $\epsilon_v\sim1$ at the inflation end)
\be
\rho_{\rm end}\sim \frac{3}{2}V(\phi_{\rm end})=\frac{3\,\lambda\,M_P^4}{2\,\alpha_1^4} \left(\frac{n}{n+\sqrt{3\alpha}}\right)^{n}\,.
\ee


\section{Reheating temperature: evaporation during PBH domination}\label{appendixB}


Upon the Universe reaching PBH domination, the reheating temperature is solely determined by the mass of the PBH at formation, and hence, is independent of both the $\beta$ parameter and the specific evolutionary trajectory that led to PBH domination. During this period, the Hubble parameter behaves as $H=2/3t$. At the point of evaporation, the temperature reaches the aforementioned reheating temperature.
\be\label{reheatpbh}
T_{\rm RH}=T_{\rm ev}=\left(\frac{40}{\pi^2}\frac{M_{P}^2}{g_*(T_{\rm RH})\,t_{\rm ev}^2}\right)^{1/4}\,,
\ee
where $t_{\rm ev}$ is the time scale associated with the evaporation point. The evaporation time scale $t_{\rm ev}$ can be estimated from the PBH mass evolution which takes the following form:
\be\label{PBHtimeev}
M=M_{\rm in}\,\left[1-\frac{\pi\,g_*(T_{\rm BH})}{160}\,\frac{M_{ P}^4}{M_{\rm in}^3}\,\left(t-t_{\rm in}\right)\right]^{\frac{1}{3}}\,.
\ee
Thus, the lifetime of the PBH is given by:
\be\label{PBHlifetime}
t_{ev}-t_{in} \sim t_{ev}=\frac{160}{\pi\,g_*(T_{\rm BH})}\,\frac{M_{\rm in}^3}{M_{P}^4}\, . 
\ee
Plugging Eq.(\ref{PBHlifetime}) into Eq.(\ref{reheatpbh}) gives
\be\label{pbhreheatingtime}
T_{\rm RH}\sim\left(\frac{g_*(T_{\rm BH})\,M_P^{10}}{640\,M_{\rm in}^6}\right)^{1/4}\,.
\ee
As we already mentioned, the expression of Eq.(\ref{pbhreheatingtime}) clearly indicates that once PBH domination is achieved, the reheating temperature depends only on the initial mass at formation.\\
The maximum allowed formation mass of PBHs $M_{\rm in}^{\rm max}$ can be calculated equating $T_{\rm ev} (T_{\rm RH})$ with BBN energy scale $T_{\rm BBN}\sim 5$ MeV
\be \label{minmax}
M_{\rm in}^{\rm max}=\left(\frac{g_*(T_{\rm BH})}{640}\right)^{\frac{1}{6}}\left(\frac{M_P^5}{T_{\rm BBN}^2}\right)^{\frac{1}{3}}\sim 2\times 10^{8}\, \rm g
\ee

\bibliographystyle{apsrev4-1}
\bibliography{main}

\end{document}